%% file: AccPars_Arxiv.tex
\documentclass[10pt,a4paper]{article}
\pdfoutput=1
\usepackage[dvipsnames]{xcolor}
\usepackage{amsmath,fourier,amssymb,amsthm,graphicx,epstopdf,tikz,hyperref,
	color,cases,float,standalone}
\usetikzlibrary{shapes,snakes}
\usetikzlibrary{calc,backgrounds}
\usepackage{chngcntr}
\counterwithout{figure}{section}
\usepackage[T1]{fontenc}
\allowdisplaybreaks
\numberwithin{equation}{section}

\theoremstyle{plain}
\newtheorem {hypo}{\bf\hspace{-\parindent}Hypothesis}[section]

\newtheorem {conj}[hypo]{Conjecture} 

\theoremstyle{remark}

\newcommand{\pf}{\begin{bpf}}

	\newcommand{\pfms}{\begin{bpfms}}
		\newcommand{\epf}{\end{bpf}\hfill$\square$\vspace{0.1cm}}
	\newcommand{\epfms}{\end{bpfms}\hfill$\square$\\ }
\newcommand\ben{\begin{equation*}}
\newcommand\ebn{\end{equation*}}
\newcommand\beq{\begin{equation}}
\newcommand\eeq{\end{equation}}
\newcommand\ds{\displaystyle}
\newcommand\lb{\left(}
\newcommand\rb{\right)}

\begin{document}

	\LARGE
	\noindent
	\textbf{Accessory parameters in confluent Heun equations \\ and classical irregular conformal blocks}
	\normalsize
	\vspace{1cm}\\
	\textit{ 
		O. Lisovyy$\,^{a}$\footnote{lisovyi@lmpt.univ-tours.fr}, A. Naidiuk$\,^{a}$\footnote{andrii.naidiuk@lmpt.univ-tours.fr}}
	\vspace{0.2cm}\\
	$^a$  Institut Denis-Poisson, Universit\'e de Tours, CNRS, Parc de Grandmont,
	37200 Tours, France
	
	\begin{abstract}
	Classical Virasoro conformal blocks are believed to be directly related to accessory parameters of Floquet type in the Heun equation and some of its confluent versions. We extend this relation  to another class of accessory parameter functions that are defined by inverting all-order Bohr-Sommerfeld periods for confluent and biconfluent Heun equation. The relevant conformal blocks involve Nagoya irregular vertex operators of rank 1 and 2 and conjecturally correspond to partition functions of a 4D $\mathcal N=2$, $N_f=3$ gauge theory at strong coupling and an Argyres-Douglas theory.
	\end{abstract}

	\section{Introduction and summary of results}
	\subsection{Heun accessory parameter and classical conformal blocks}
	The Heun's differential equation (HE), written in normal form, is given by 
    \beq \label{heune}
	\begin{gathered}
	\psi''\lb z\rb=V\lb z\rb \psi\lb z\rb,\\ V\lb z\rb=\frac{\theta_0^2-\frac14}{z^2}+\frac{\theta_1^2-\frac14}{\lb z-1\rb^2}+\frac{\theta_t^2-\frac14}{\lb z-t\rb^2}+\frac{\theta_\infty^2-\theta_0^2-\theta_1^2-\theta_t^2+\frac12}{z\lb z-1\rb}+
	\frac{\lb 1-t\rb \mathcal E}{z\lb z-1\rb\lb z-t\rb}.
	\end{gathered}
	\eeq
    Every 2nd order ODE with at most 4 regular singular points  on the Riemann sphere can be reduced to \eqref{heune} by a M\"obius transformation mapping these points to $0,t,1,\infty$.  The exponents of local monodromy are encoded into the Riemann scheme
	\begin{center}
	\begin{tabular}{c|c|c|c}
	0 & 1 & t & $\infty$ \\  \hline \rule{0pt}{2.5ex}
	$\frac12\pm\theta_0$ & $\frac12\pm\theta_1$ & $\frac12\pm\theta_t$ & $\frac12\pm\theta_\infty$	
	\end{tabular}	
	\end{center}
    The main new feature of \eqref{heune}, as compared to the case of 3 regular singularities described by the Gauss hypergeometric equation, is the presence of an \textit{accessory parameter}~$\mathcal E$, which does not influence the local behavior of solutions. This makes the global analysis of Heun's equation considerably more difficult, the central question being the dependence of  $\mathcal E$ on monodromy.
    
    Indeed, the Riemann-Hilbert correspondence assigns to every choice of $\mathcal E$ and $t$ a point in the space of monodromy data which consists of conjugacy classes of $4$-tuples of $\mathrm{SL}\lb 2,\mathbb C\rb$-matrices with fixed spectrum that multiply to identity:
    \beq
    \mathcal M=\bigl\{M_0,M_1,M_t,M_\infty\in\mathrm{SL}\lb 2,\mathbb C\rb: M_\infty M_1 M_t M_0=\mathbf 1,\operatorname{Tr}M_k=-2\cos2\pi \theta_k\text{ for }k=0,1,t,\infty\bigr\}\bigl/\sim.
    \eeq
    The space $\mathcal M$ is generically 2-dimensional and  $\lb \mathcal E,t\rb$ can therefore be regarded as a pair of local coordinates on it. Any other convenient coordinate $\xi$ on $\mathcal M$ may be viewed as a function of $\lb \mathcal E,t\rb$. Inverting this relation, to any choice of $\xi$ we may assign an accessory parameter function $\mathcal E^{[\xi]}\lb t|\,\xi\rb$, where the upper index is introduced to emphasize that $\mathcal E^{[\xi]}$ depends not only on the value of $\xi$ but also on how this coordinate was chosen. A standard choice is to take as $\xi$ one of the three Floquet exponents $\sigma_{kt}$ defined by 
    \beq
    \operatorname{Tr}M_kM_t=-2\cos2\pi\sigma_{kt},\qquad k=0,1,\infty.
    \eeq
    We thereby obtain three different accessory parameter functions --- however, since different singular points of \eqref{heune} can be exchanged by M\"obius transformations, these functions turn out to be related by a permutation of parameters $\vec\theta=\lb\theta_0,\theta_1,\theta_t,\theta_\infty\rb$ combined with an appropriate transformation of $t$.
    
    The \textit{direct} monodromy problem for Heun's equation consists essentially in finding $\sigma_{kt}$'s for given  $\lb \mathcal E,t\rb$; it can be formally solved using Hill determinants. Remarkably, the \textit{inverse} monodromy problem of reconstruction of Heun's equation with prescribed monodromy exponent $\sigma_{kt}$ can be solved more efficiently.  The result is given by a perturbative series such as
    \beq
    \mathcal E^{[\sigma_{0t}]}\lb t|\,\sigma\rb=\sum_{n=0}^{\infty}\mathcal E^{[\sigma_{0t}]}_n\lb \sigma\rb t^n,
    \eeq
    whose coefficients $\mathcal E^{[\sigma_{0t}]}_k$ can be found recursively using continued fractions.
    Both Hill determinant and continued fraction method are well-known since the end of 19th century in the context of Mathieu equation, see e.g. \cite[Chapter 19]{WW}, \cite[Chapter 20]{AS}, \cite{Heine}, \cite[\href{http://dlmf.nist.gov/28.15.E1}{Eq.28.15.E1}]{DLMF}. The Mathieu counterparts of functions $\sigma_{0t}\lb\mathcal E,t\rb$ and $\mathcal E^{[\sigma_{0t}]}\lb t|\,\sigma\rb$ are implemented in Mathematica as \texttt{Mathieu\-Charac\-teristicExponent} and \texttt{MathieuCharacteristicA}. Although extension of continued fractions to Heun is straightforward, it does not seem to be widely known and is included to Appendix~\ref{appendixHeun}.
    
    The second class of special functions studied in this work are Virasoro conformal blocks (CBs). The regular 4-point spherical  $s$-channel CB, 
    \beq\label{cblock}
    \mathcal F\lb t|c,\left\{\Delta_k\right\}\rb =t^{\Delta_\sigma-\Delta_0-\Delta_t}\left[1+\sum_{n=1}^{\infty}\mathcal F_n\lb c,\left\{\Delta_k\right\}\rb t^n\right].
    \eeq
    depends on the anharmonic ratio $t$ and six complex parameters: the Virasoro central charge $c$ and five conformal dimensions $\Delta_{0,1,t,\infty,\sigma}$. The coefficients $\mathcal F_n$ are rational in $c,\left\{\Delta_k\right\}$ and are fixed by the commutation relations of the Virasoro algebra \cite{BPZ}. Their explicit form can also be obtained using the AGT correspondence \cite{AGT} identifying $\mathcal F\lb t|c,\left\{\Delta_k\right\}\rb$ with Nekrasov partition function \cite{Nekrasov,NO} of $\mathcal N=2$ 4D supersymmetric $\mathrm{SU}\lb 2\rb$ gauge theory with $N_f=4$ flavors. 
    
    The relation between Heun accessory parameter functions and regular CBs is given by a composition of two conjectures
    formulated in 1986 paper of Zamolodchikov \cite[Eqs. (2.30)--(2.32)]{Zamo86}\footnote{\cite{Zamo86} further refers to \cite{BPZ}; however, the discussion of the quasiclassical limit in \cite{BPZ} is limited to a footnote on p.~357 which does not contain a neat formulation of any of the two hypotheses.}:
     \begin{itemize}
    	\item \textit{Conjecture A (exponentiation)}. It states that in the quasiclassical limit
    	\beq\label{qclimit1}
    	c,\Delta_{0,1,t,\infty,\sigma}\to\infty,\qquad \frac{6\Delta_k}{c}\to\delta_k,\qquad \qquad k=0,1,t,\infty,\sigma
    	\eeq
    	conformal block \eqref{cblock} has WKB type asymptotics 
    	\beq\label{qclimit2}
    	\mathcal{F}\lb t\,\bigl|\left\{\Delta_k\right\}\rb\sim \exp  \frac{c}{6}\mathcal{W}\lb t\,\bigl|\left\{\delta_k\right\}\rb.
    	\eeq 
    	The series
    	\beq\label{clcblock}
    	\mathcal{W}\lb t\,\bigl|\left\{\delta_k\right\}\rb=\lb \delta_\sigma-\delta_0-\delta_t\rb\ln t +
    	\sum_{n=1}^\infty \mathcal{W}_n\lb \left\{\delta_k\right\}\rb t^n
    	\eeq
    	appearing in this asymptotics is called \textit{classical conformal block}. 
    	\item \textit{Conjecture B (Heun/classical CB correspondence)}. Classical CB and Heun accessory parameter function  are related by
    	\beq\label{polyacon}
    	\mathcal E^{[\sigma_{0t}]}\bigl( t|\,\sigma,\vec\theta\bigr)=t\frac{\partial}{\partial t}\mathcal{W}\lb t\,\bigl|\left\{\delta_k\right\}\rb.
    	\eeq
    	where the rescaled conformal dimensions are identified with Heun monodromy exponents via
    	\beq\label{scdims}
    	\delta_\sigma=\tfrac14-\sigma^2,\qquad \delta_k=\tfrac14-\theta_k^2,\qquad k=0,1,t,\infty.
    	\eeq
    	    \end{itemize}
    Conjecture A is supported by semiclassical arguments in Liouville QFT and on the gauge theory side of the AGT correspondence where $\mathcal{W}\lb t\rb$ is interpreted as Nekrasov-Shatashvili effective twisted superpotential \cite{NS}. The strongest evidence comes from the explicit computation of large-order CB coefficients. We also mention a recent paper \cite{BDK} which claims to prove the exponentiation of the 4-point CB \eqref{cblock}. Conjecture~B follows from a refined version of the exponentiation (the so-called heavy-light factorization) for conformal blocks with degenerate fields combined with BPZ decoupling equations \cite{T10,LLNZ} (see also \cite{PP3} for details). It can also be tested directly by comparing the classical CB series \eqref{clcblock} order by order with the accessory parameter expansion computed from continued fractions or by other approaches \cite{NC,LN,HK,Menotti}.

    \subsection{Confluent Heun equations}
	Our aim is to extend Conjectures A and B to the irregular setting. Indeed, there is a number of confluent versions of the Heun's equation \cite{Slavyanov} which we organize into a geometric degeneration diagram:
	
    
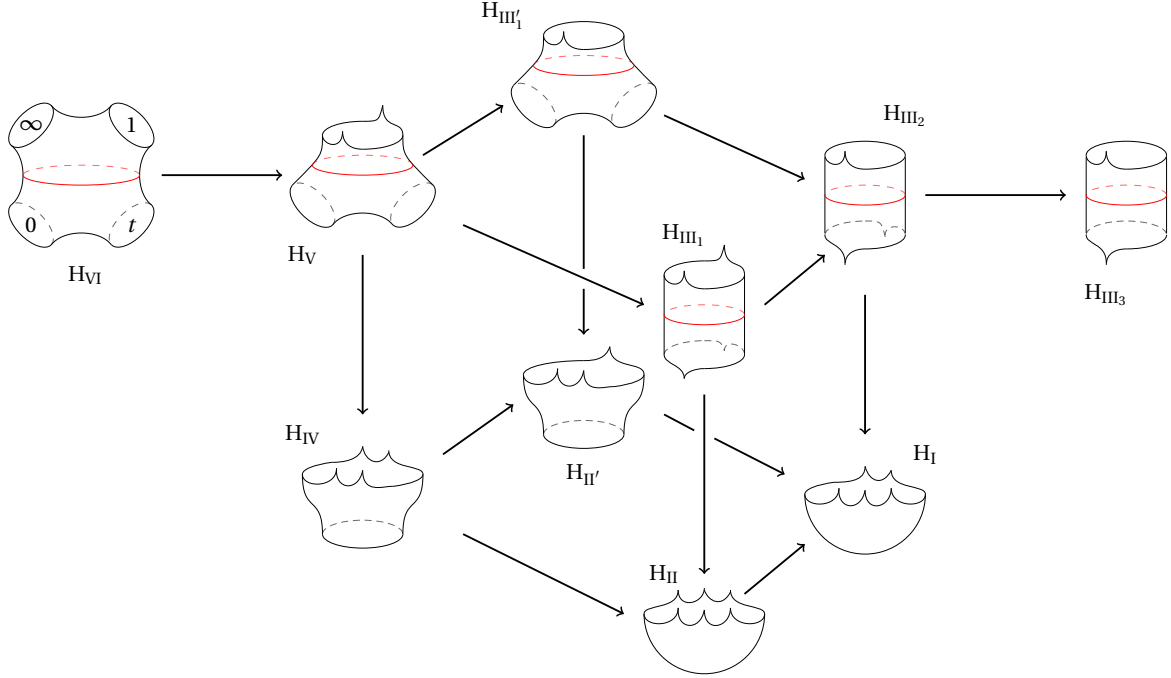
\begin{figure}[H]
    	\centering
	\includestandalone[height=9cm]{diagram}
   \end{figure}
	
   \noindent Each entry of this diagram represents an ODE of the form $\psi''\lb z\rb=V\lb z\rb\psi\lb z\rb$ on $\mathbb{CP}^1$. The form of the potential $V\lb z\rb$ can be read off from the corresponding Riemann surface. Every hole corresponds to a pole of the quadratic differential $V\lb z\rb dz^2$ whose order is equal to the number of cusps plus 2, so that non-cusped hole corresponds to a regular singular point. Thus H$_\mathrm{VI}$ is the usual HE, H$_\mathrm{V}$ has two regular singular points and an irregular one, etc. We record the potentials in the following table:
   	\begin{table}[H]
   		   \begin{center}
   \begin{tabular}{c|c|c}
   notation & equation name & potential $V\lb z\rb$ \\ \hline\hline
   \rule{0pt}{3.5ex} \rule[-1.5ex]{0pt}{0pt}
   H$_\mathrm{V}$ & confluent HE  & $\frac{\theta_0^2-\frac14}{z^2}+\frac{\theta_t^2-\frac14}{\lb z-t\rb^2}+\frac14+\frac{\theta_*}{z}-\frac{\mathcal E}{z\lb z-t\rb}$ \\ 
   \hline  \rule{0pt}{3.5ex} \rule[-2ex]{0pt}{0pt}H$_\mathrm{IV}$ &  biconfluent HE &  $\frac{\theta_0^2-\frac14}{z^2}-\frac{\mathcal E}{z}+2\theta_{\bullet}+\lb z+t\rb^2$ \\
   \hline 
   \rule{0pt}{3.5ex} \rule[-1.5ex]{0pt}{0pt}  H$_\mathrm{III_1}$ & doubly confluent HE & $\frac{t^2}{4z^4}+\frac{t\theta_\star}{z^3}-\frac{\mathcal E}{z^2}+\frac{\theta_*}{z}+\frac14$ \\
   \hline 
   \rule{0pt}{3.5ex} \rule[-1.5ex]{0pt}{0pt}
   H$_\mathrm{III_2}$ & reduced doubly confluent HE & $\frac{t}{z^3}-\frac{\mathcal E}{z^2}+\frac{\theta_*}{z}+\frac14$ \\
   \hline 
   \rule{0pt}{3.5ex} \rule[-1.5ex]{0pt}{0pt}
   H$_\mathrm{III_3}$ & doubly reduced doubly confluent HE & $\frac{t}{z^3}-\frac{\mathcal E}{z^2}+\frac{1}{z}$ \\
   \hline
   \rule{0pt}{3.5ex} \rule[-1.5ex]{0pt}{0pt}
    H$_\mathrm{II}$ & triconfluent HE & $\lb z^2+t\rb^2+2\theta_\circ z +\mathcal E$ \\
   \hline
   \rule{0pt}{3.5ex} \rule[-1.5ex]{0pt}{0pt}
    H$_\mathrm{I}$ & reduced triconfluent HE & $4z^3+2tz+\mathcal E$ \\ 
   \hline\hline
   \rule{0pt}{3.5ex} \rule[-1.5ex]{0pt}{0pt}  H$_\mathrm{III_1'}$ & reduced confluent HE &
   $\frac{\theta_0^2-\frac14}{z^2}+\frac{\theta_t^2-\frac14}{\lb z-t\rb^2}+\frac{1}{z}-\frac{\mathcal E}{z\lb z-t\rb}$
   \\ \hline
   \rule{0pt}{3.5ex} \rule[-1.5ex]{0pt}{0pt}  H$_\mathrm{II'}$ & reduced biconfluent HE & $\frac{\theta_0^2-\frac14}{z^2}+\frac{\mathcal E}{z}+t+z$
   \end{tabular}
   \end{center} 
   \caption{Notation for confluent Heun equations}
   \label{Table1}
   \end{table} 
   The parameters such as $\theta_0,\theta_t,\theta_*,\theta_\bullet,\theta_\star,\theta_\circ$ are exponents of local monodromy around regular singular points or formal monodromy around the irregular ones; $\mathcal E$ denotes the accessory parameter. The ``time'' (or ``coupling'') $t$ parameterizes the exponential behavior of $\psi\lb z\rb$ at irregular singularities, except in the case of H$_\mathrm{V}$ and  H$_\mathrm{III_1'}$,  where it can be equivalently chosen as the position of one of the 2 regular singular points.

   A Laplace transform  of H$_\mathrm{III_1}$ and H$_\mathrm{II}$ (applied to their canonical forms) transforms them into H$_\mathrm{III_1'}$ and H$_\mathrm{II'}$. The above scheme thus contains seven inequivalent confluent HEs. It is not a surprise that the geometric confluence diagram exactly reproduces the one suggested in \cite{CM,CMR} for Lax pairs of Painlevé equations. Indeed, there exists a connection between Heun and Painlevé functions \cite[Section 2.1.3 and Chapter 15]{FIKN}, \cite{Nov86} going back to classical papers of Fuchs \cite{Fuchs} and Garnier \cite{Garnier} which was also the subject of one of the last works of Boris Dubrovin \cite{DK}.
   
   Confluent HEs appear in a wide range of physical applications. To mention a few examples, H$_\mathrm{I}$ and H$_\mathrm{II}$ are Schroedinger equations for cubic and quartic oscillator; H$_\mathrm{III_3}$ is equivalent to the Mathieu equation (quantum pendulum) through the change of variables $f\lb x\rb=e^{-ix}\psi\lb \sqrt t\,e^{2ix}\rb$; H$_\mathrm{V}$ arises in the black hole scattering \cite{BPM,Lea,STU,Fiziev,CCN} and Rabi model of quantum optics \cite{MPS,ZXBL,CCR}, etc.
   
   \subsection{Conjectures A and B: confluent case}
   There already exist a few irregular extensions of Conjectures A and B. Immediately after the AGT discovery of 4D/2D duality it was realized that Nekrasov partition functions of $\mathcal N=2$ SUSY gauge theories with lower number of flavors ($N_f=0,1,2,3$)  correspond to a class of irregular CBs; we will call them \textit{confluent CBs of the 1st kind}. The algebraic construction of these CBs \cite{Gaiotto09,BMT,GT} involves rank~1 Whittaker modules for the Virasoro algebra. They are  given by \textit{weak coupling} ($t\to0$) expansions similar to \eqref{cblock} whose coefficients are nothing but suitable limits of $\mathcal F_n\lb c,\left\{\Delta_k\right\}\rb$. The same applies to quasiclassical limit where confluent CBs of the 1st kind are expected to exponentiate \cite{PP1,RZ1,RZ2,PP2} and produce Nekrasov-Shatashvili twisted superpotentials  of the relevant gauge theories.
   
   On the other hand, for confluent equations H$_\mathrm{V}$, H$_{\mathrm{III}_1}$, H$_{\mathrm{III}_2}$, H$_{\mathrm{III}_3}$ there exists a natural counterpart of the accessory parameter function $\mathcal E^{[\sigma_{0t}]}\lb t|\,\sigma\rb$. Recall that the latter is fixed by the choice of monodromy parameter~$\sigma_{0t}$, which in the confluent cases should be replaced by the Floquet exponent of (non-formal!) monodromy around irregular singular point at~$\infty$. The relevant cycles are shown in red color in the degeneration diagram. Slightly abusing the notation, we will denote the corresponding confluent accessory parameter function by $\mathcal E^{[\mathrm{F}]}\lb t|\,\sigma\rb$ and call it \textit{Floquet characteristic}. Its small $t$ expansion can be computed using continued fractions exactly as in the Heun case (see Appendix~\ref{appCH}). We also have
   \begin{subequations}
   	\label{confchain}
   \begin{align}
   \label{conflimit1}\mathcal E^{[\mathrm{F}]}_{\mathrm{V}}\lb t|\,\sigma\rb=&\,\lim_{\Lambda\to\infty}\mathcal E^{[\mathrm{F}]}_{\mathrm{VI}}\lb \tfrac{t}{\Lambda}|\,\sigma\rb\Bigl|_{\theta_{1}=\frac{\Lambda+\theta_*}{2},\theta_{\infty}=\frac{\Lambda-\theta_*}{2}},\\
   \mathcal E^{[\mathrm{F}]}_{\mathrm{III}_1}\lb t|\,\sigma\rb=&\,\lim_{\Lambda\to\infty}\biggl[\mathcal E^{[\mathrm{F}]}_{\mathrm{V}}\lb \tfrac{t}{\Lambda}|\,\sigma\rb-\theta_0^2-\theta_t^2+\frac12\biggr]_{\theta_{0}=\frac{\Lambda-\theta_\star}{2},\,\theta_{t}\,=\frac{\Lambda+\theta_\star}{2}},\\
   \mathcal E^{[\mathrm{F}]}_{\mathrm{III}_2}\lb t|\,\sigma\rb=&\,\lim_{\theta_\star\to\infty}\mathcal E^{[\mathrm{F}]}_{\mathrm{III}_1}\lb t/\theta_\star|\,\sigma\rb,\\
   \mathcal E^{[\mathrm{F}]}_{\mathrm{III}_3}\lb t|\,\sigma\rb=&\,\lim_{\theta_*\to\infty}\mathcal E^{[\mathrm{F}]}_{\mathrm{III}_2}\lb t/\theta_*|\,\sigma\rb.
   \end{align}
   \end{subequations} 

   A relation between the Floquet characteristic $\mathcal E^{[\mathrm{F}]}_{\mathrm{III}_3}\lb t|\,\sigma\rb$ (i.e. accessory parameter in the Mathieu equation) with Nekrasov-Shatashvili twisted superpotential $\mathcal W_{N_f=0}\lb t\rb$ of the pure  gauge theory has appeared in \cite{NS} as the simplest example of Bethe/gauge correspondence. Using the AGT relation, this correspondence was reformulated 
   in terms of classical irregular CBs in \cite{PP1}. Further extensions to $N_f=1,2$ in \cite{PP2,RZ1} and $N_f=3$ in \cite{CCC} similarly involve Floquet characteristics of confluent equations H$_\mathrm{III_2}$, H$_\mathrm{III_1}$ and H$_\mathrm{V}$. In all four cases, the analog of Conjecture B has the same form 
   as \eqref{polyacon}:
   \beq\label{aux1823}
   \mathcal E^{[\mathrm{F}]}\lb t|\,\sigma\rb=t\frac{\partial}{\partial t}\mathcal W_{N_f}\lb t\rb.
   \eeq
   These developments are briefly reviewed in Subsection~\ref{secconfCB1} and Appendix~\ref{appCH} for convenience of the readers with little prior exposition to 4D/2D vocabulary, and also to provide general statements in a uniform notation\footnote{For $N_f=2$, Refs. \cite{PP2,RZ1} deal (on the differential equations side) with the Whittaker-Hill equation which is a special case of H$_\mathrm{III_1}$.  Ref. \cite{CCC} uses gauge theory quantities to define the analog of the right side of \eqref{aux1823} for $N_f=3$.}.
   
   The main goal of the present work is to initiate the study of confluent counterparts of Conjectures~A and B in the \textit{strong coupling} ($t\to\infty$) regime. In this setting, it is much less clear what one should expect at both sides of the putative analog of the relation \eqref{polyacon}. First, one needs to introduce a new basis in the space of irregular CBs to describe the confluent limit of the regular CBs in the $u$-channel. Several candidates for these \textit{confluent CBs of the 2nd kind} were suggested in the work of Gaiotto and Teschner \cite{GT} which defines a collision limit of the regular $u$-channel CBs leading to finite answers. An algebraic definition of such CBs was proposed by Nagoya in \cite{Nagoya1,Nagoya2}; its main new feature is an irregular vertex operator acting between two Virasoro-Whittaker modules.
   
    Note that the series for confluent CBs of the 2nd kind cannot be expected to converge. Instead, they should be considered as formal series in $\frac1t$ which represent asymptotic expansions of an actual function inside some sectors (or even only along some directions) at $\infty$. Different sectors/directions can in principle lead to different bases of confluent CBs of the 2nd kind and it may well turn out that the constructions of \cite{GT,Nagoya1,Nagoya2} do not cover all of them. In any case, it is straightforward to test the quasiclassical asymptotics for confluent CBs of the 2nd kind that are already available. We address this question in Subsection~\ref{subsec2kind} by checking exponentiation  (Conjecture A) in two cases: 
    (a) 3-point CB  with 2 regular punctures and 1 irregular puncture of rank 1 (b) 2-point CB with 1 regular and 1 irregular puncture of rank 2.

   Trying to define the accessory parameter side of the confluent Conjecture B for large $t$, one faces a conceptual problem of ``good'' choice of the monodromy parameter defining the accessory parameter function, as well as more technical issue of its efficient computation.  We propose to replace the Floquet exponent $\sigma$ by the all-order Bohr-Sommerfeld (BS) period
   \beq
   \nu=\frac{1}{2\pi i}\oint_\gamma S_{\mathrm{odd}}\lb z\rb dz,\qquad S_{\mathrm{odd}}\lb z\rb=\sum_{n=0}^\infty S_{2n-1}\lb z\rb,
   \eeq
   for a suitably chosen closed contour $\gamma$, with $S_{-1}\lb z\rb=\ds\sqrt{\ds V\lb z\rb}$ and $S_n'+\sum_{k=-1}^{n+1}S_{k}S_{n-k}=0$. The computation of BS periods is a standard tool of $\mathcal N=2$ SUSY gauge theory \cite{MM,HM}; see also \cite{BD,GGM} and references therein. In the gauge theory context, the role of the Planck constant is played by one of the Nekrasov deformation parameters $\epsilon_{1,2}$ (the other one being 0 in the NS limit). In our setup, however, the relevant WKB parameter is the inverse coupling $\tfrac1t$. 
   
   We focus on two examples: (a) confluent and (b) biconfluent Heun equations H$_\mathrm{V}$ and H$_\mathrm{IV}$. Inverting the relation $\nu=\nu\lb\mathcal E,t\rb$, it is fairly easy to obtain accessory parameter $\mathcal E=\mathcal E^{\text{[BS]}}\lb t\,|\,\nu\rb$ in the form of a formal asymptotic series in $\tfrac1t$. These series are then identified with classical confluent CBs via Conjectures~\ref{conjHVCFT} and~\ref{conjHIVCFT}. We find that, similarly  to \eqref{polyacon} and \eqref{aux1823}, 
   \beq
  \mathcal E^{\text{[BS]}}_{\mathrm{V}}\lb t\,|\,\nu\rb=t\frac{\partial}{\partial t}\mathcal U_{N_f=3}\lb t\rb,\qquad 
  \mathcal E^{\text{[BS]}}_{\mathrm{IV}}\lb t\,|\,\nu\rb=\frac{\partial}{\partial t}\tilde{\mathcal U}\lb t\rb,
   \eeq
   where $\mathcal U_{N_f=3}\lb t\rb$ and $\tilde{\mathcal U}\lb t\rb$ are classical counterparts of CBs introduced in \cite{GT,Nagoya1}. This observation is consistent with a conjectural relation between conformal blocks of the 2nd kind and partition functions of strongly coupled gauge/Argyres-Douglas theories \cite{GT,BLMST,NU}, and may be viewed as a quasiclassical variant of the AGT or Bethe/gauge correspondence.

    \section{Classical confluent conformal blocks}
    \subsection{Regular conformal blocks and their classical limit}
    Regular CBs are matrix elements of compositions of vertex operators between states in the highest weight representations of the Virasoro algebra
    \beq 
    \left[L_m,L_n\right]=\lb m-n\rb L_{m+n}+\frac{c}{12}n\lb n^2-1\rb\delta_{m+n,0}.
    \eeq
    A highest weight representation $\mathcal V_{\Delta}$ and its dual  $\mathcal V_{\Delta}^*$ are generated from the states $|\Delta\rangle$, $\langle\Delta|$ defined by
    \beq\begin{aligned}
    &L_0|\Delta\rangle=\Delta|\Delta\rangle,\qquad L_{n>0}|\Delta=0,\\
    &\langle \Delta|L_0=\Delta\langle \Delta|,\qquad \langle \Delta|L_{n<0}=0.
    \end{aligned}
    \eeq
    There is a canonical bilinear pairing $\langle|\rangle:\mathcal V_{\Delta}^*\times \mathcal V_{\Delta}\to\mathbb C$ satisfying $ \langle u|L_n\cdot|v\rangle=\langle u|\cdot L_n|v\rangle$ for all $\langle u|\in \mathcal V_{\Delta}^*$, $|v\rangle\in \mathcal V_{\Delta}$ and $n\in \mathbb Z$. It will be normalized by $\langle \Delta|\Delta\rangle=1$. The primary regular vertex operators $V_{\Delta_3,\Delta_1}^{\Delta_2}\lb t\rb:
    \mathcal V_{\Delta_1}\to \mathcal V_{\Delta_3}$ are defined by the commutation relation
    \beq\label{VOVir}
    \left[ L_n, V_{\Delta_3, \Delta_1}^{\Delta_2}\lb t\rb
    \right]=t^n\left( t \tfrac{\partial}{\partial t}+
    \lb n+1\rb\Delta_2\right)
    V_{\Delta_3, \Delta_1}^{\Delta_2}\lb t\rb.
    \eeq
    It determines $V_{\Delta_3,\Delta_1}^{\Delta_2}\lb t\rb$ uniquely up to normalization (multiplicative constant independent of $t$), which we further fix by setting $\langle \Delta_3| V_{\Delta_3, \Delta_1}^{\Delta_2}\lb t\rb|\Delta_1\rangle=t^{\Delta_3-\Delta_2-\Delta_1}$. Spherical 4-point conformal block \eqref{cblock} is defined as
    \beq\label{4blCVO}
    \mathcal F\lb t|c,\left\{\Delta_k\right\}\rb:=\bigl\langle\Delta_\infty\bigl|V_{\Delta_\infty,\Delta_\sigma}^{\Delta_1}\lb 1\rb V_{\Delta_\sigma,\Delta_0}^{\Delta_t}\lb t\rb\bigr|\Delta_0\bigr\rangle =\vcenter{\hbox{\includestandalone[height=2cm]{4point}}}
    \eeq
    The low-order expansion coefficients in $\mathcal F\lb t|c,\left\{\Delta_k\right\}\rb=t^{\Delta_\sigma-\Delta_0-\Delta_t}\left[1+\sum_{n=1}^{\infty}\mathcal F_n\lb c,\left\{\Delta_k\right\}\rb t^n\right]$ can be computed by inserting  the resolution of the identity operator in a suitable basis of states in $\mathcal V_{\Delta_\sigma}$ between the two vertex operators in \eqref{4blCVO}. For example, one has
    \begin{subequations}
    \begin{align}
    \label{F1expr}\mathcal F_1\lb c,\left\{\Delta_k\right\}\rb=&\,\frac{\lb \Delta_\sigma-\Delta_0+\Delta_t\rb \lb \Delta_\sigma-\Delta_\infty+\Delta_1\rb}{2\Delta_\sigma}, \\
     \label{F2expr} \mathcal F_2\lb c,\left\{\Delta_k\right\}\rb=&\,\frac{\left(\Delta_\sigma-\Delta_0+\Delta_t\right)\left(\Delta_\sigma-\Delta_0+\Delta_t+1\right)
    	\left(\Delta_\sigma-\Delta_\infty+\Delta_1\right)\left(\Delta_\sigma-\Delta_\infty+\Delta_1+1\right)}{4\Delta_\sigma\left(1+2\Delta_\sigma\right)}\,+\Biggr.\\
    \nonumber\Biggl.+&\,\frac{\left(1+2\Delta_\sigma\right)\left(\Delta_0+\Delta_t+\frac{\Delta_\sigma\left(\Delta_\sigma-1\right)-
    		3\left(\Delta_0-\Delta_t\right)^2}{1+2\Delta_\sigma}\right)\left(\Delta_\infty+\Delta_1+\frac{\Delta_\sigma\left(\Delta_\sigma-1\right)-
    		3\left(\Delta_\infty-\Delta_1\right)^2}{1+2\Delta_\sigma}\right)}{
    	2\left(1-4\Delta_\sigma\right)^2+2\left(c-1\right)\left(1+2\Delta_\sigma\right)}.
    \end{align}
    \end{subequations}
   	The form of $\mathcal F_n$'s becomes quite involved with the growth of $n$ but all coefficients can still be explicitly computed for arbitrary $n$ thanks to the AGT relation. For concrete formulas, see e.g.	\cite{AFLT} (or \cite[Eqs. (1.2)--(1.4)]{LNR}, whose notation is close to the present paper).
   	
   	The quasiclassical limit described by \eqref{qclimit1}--\eqref{clcblock} more precisely means that one should consider $\frac{6}{c}\ln \mathcal F\lb t|c,\left\{\Delta_k\right\}\rb$ as a formal series in $t$, and every coefficient of this series has a finite limit (Conjecture A). E.g. from \eqref{F1expr}--\eqref{F2expr} we find
   	    \begin{subequations}
   		\begin{align}
   		\label{W1expr}\mathcal W_1\lb \left\{\delta_k\right\}\rb=&\,\frac{\lb \delta_\sigma-\delta_0+\delta_t\rb \lb \delta_\sigma-\delta_\infty+\delta_1\rb}{2\delta_\sigma}, \\
   		\label{W2expr} \mathcal W_2\lb \left\{\delta_k\right\}\rb=&\,
   		\frac{\lb \delta_\sigma-\delta_0+\delta_t\rb^2 \lb \delta_\sigma-\delta_\infty+\delta_1\rb^2}{8\delta_\sigma^2}\left[\frac{1}{\delta_\sigma-\delta_0+\delta_t}+\frac{1}{\delta_\sigma-\delta_\infty+\delta_1}-\frac{1}{2\delta_\sigma}\right]+\\
   		\nonumber+&\,\frac{\lb\delta_\sigma^2+2\delta_\sigma\lb\delta_0+\delta_t\rb-3\lb\delta_0-\delta_t\rb^2\rb \lb\delta_\sigma^2+2\delta_\sigma\lb\delta_\infty+\delta_1\rb-3\lb\delta_\infty-\delta_1\rb^2\rb}{16\delta_\sigma^2\lb 4\delta_\sigma+3\rb}.
   		  \end{align}
   	\end{subequations}
   	Already at the order $O\lb t^2\rb$, the existence of the limit is not immediately obvious and involves a cancellation of separately divergent contributions from $\mathcal F_1$ and $\mathcal F_2$. 

    In the discussion of confluent limits, it will be occasionally convenient to use the Liouville CFT parameterization of the Virasoro central charge and conformal dimensions, 
    \beq
    c=1+6\lb b+b^{-1}\rb^2,\qquad \Delta=\frac{c-1}{24}+P^2. 
    \eeq
    The classical limit of the regular CB corresponds to setting
    $P_k= ib^{-1}\theta_k $  ($k=0,1,t,\infty,\sigma$) and  sending $b$ to $0$, so that $c\sim 6b^{-2}$ and ${\Delta_k\sim b^{-2}\delta_k}$ with $\delta_k=\frac14-\theta_k^2$. When it will be useful to indicate the dependence of CBs such as $\mathcal F\lb t|c,\left\{\Delta_k\right\}\rb$ and $\mathcal W\lb t|\left\{\delta_k\right\}\rb$ on various conformal dimensions more explicitly, we will accordingly use a notation such as $\mathcal F\lb\substack{P_{1}\;\quad 
    	P_{t}\\ P_\sigma \\ P_{\infty}\quad P_0};t\rb$ and $\mathcal W\lb\substack{\theta_{1}\;\quad 
    	\theta_{t}\\ \sigma \\ \theta_{\infty}\quad \theta_0};t\rb$.
    
    \subsection{Confluent conformal blocks of the 1st kind \label{secconfCB1}}
    \subsubsection{Whittaker modules}
    In the confluent case, in addition to the highest weight representations for the Virasoro algebra, one also needs to consider Whittaker modules. A Whittaker module $\mathcal{V}^{W}_{\vec\lambda}$ of rank $r$ is generated from a joint eigenstate $\bigl|\vec{\lambda}\bigr\rangle=\left|\lambda_r,\ldots,\lambda_{2r}\right\rangle$ of Virasoro generators $L_r,\ldots,L_{2r}$ \cite{Gaiotto09,BMT,GT}:
    \beq
    L_n\left|\lambda_r,\ldots,\lambda_{2r}\right\rangle=\begin{cases}
    	\lambda_n\left|\lambda_r,\ldots,\lambda_{2r}\right\rangle,\qquad & r\leq n\leq 2r,\\
    	\quad 0,\qquad & n>2r,
    \end{cases}
    \eeq
    which is called a Whittaker vector.
    Note that $e^{s L_n}\bigl|\vec{\lambda}\bigr\rangle$ with $n=0,\ldots, r-1$ and $s\in\mathbb C$ is again a Whittaker vector. For example,
    \begin{align}
    t^{L_0}\left|\lambda_r,\ldots,\lambda_{2r}\right\rangle\sim&\, \left|t^r\lambda_r,\ldots,t^{2r}\lambda_{2r}\right\rangle,
    \end{align}
    This property of  $e^{s L_n}\bigl|\vec{\lambda}\bigr\rangle$ will be used later to fix some of the eigenvalues in the definitions of confluent CBs. 
   The present subsection is concerned only with Whittaker modules of rank 1. In this case, there is a canonical bilinear pairing $\langle|\rangle:\mathcal{V}^{W,*}_{\lb\lambda_1,\lambda_2\rb}\times \mathcal V_{\Delta}\to\mathbb C$ which is uniquely determined by the condition
    $\langle w|\cdot L_n|v\rangle=\langle w| L_n \cdot|v\rangle$ for all $\langle w|\in
    \mathcal{V}^{W,*}_{\lb\lambda_1,\lambda_2\rb}$, $|v\rangle\in \mathcal V_{\Delta}$ and $n\in \mathbb Z$,
    and normalization $\langle \lambda_1,\lambda_2|\Delta\rangle=1$.

    \subsubsection{$N_f=3$}
    The confluent CB of the 1st kind, AGT-dual to the $N_f=3$ $\mathrm{SU}\lb 2\rb$ Nekrasov partition function, can be defined as the matrix element
    \beq\label{confCBNf3}
    \mathcal F_{N_f=3}\lb t\,\bigl|\,c;P_0,P_t,P_*,P_\sigma\rb:=\left\langle P_*,\tfrac14\bigl|V_{\Delta_\sigma,\Delta_0}^{\Delta_t}\lb t\rb\bigr|\Delta_0\right\rangle
    =\vcenter{\hbox{\includestandalone[height=2cm]{conflBlock1}}}
    \eeq
    Its small $t$ expansion may be found by expanding $V_{\Delta_\sigma,\Delta_0}^{\Delta_t}\lb t\rb\bigr|\Delta_0\bigr\rangle$ in a suitable basis of $\mathcal V_{\Delta_\sigma}$ and computing the pairing with $\bigl\langle P_*,\tfrac14\bigl|$. The first few coefficients are explicitly given by
    \begin{align}
    \label{clCBNf3}
    &\mathcal F_{N_f=3}\lb t\,\bigl|\,c;\{P_k\}\rb=t^{\Delta_\sigma-\Delta_0-\Delta_t}\Biggl[1+\frac{\lb \Delta_\sigma-\Delta_0+\Delta_t\rb P_*}{2\Delta_\sigma}t+\\
    \nonumber +&\,\Biggl(\frac{\left(\Delta_\sigma-\Delta_0+\Delta_t\right)\left(\Delta_\sigma-\Delta_0+\Delta_t+1\right)P_*^2}{\Delta_\sigma\left(1+2\Delta_\sigma\right)}+
    \frac{\left(\Delta_0+\Delta_t+\frac{\Delta_\sigma\left(\Delta_\sigma-1\right)-
    		3\left(\Delta_0-\Delta_t\right)^2}{1+2\Delta_\sigma}\right)\left(1+2\Delta_\sigma-6P_*^2\right)}{16\Delta_\sigma^2-10\Delta_\sigma+c\left(1+2\Delta_\sigma\right)}\Biggr)\,\frac{ t^2}{4}+O\lb t^3\rb\Biggr].
    \end{align}
    This series can also be obtained by taking termwise confluent limit of the expansion \eqref{cblock} where 2 of the 5 conformal dimensions become infinite in a correlated way:
    \beq
    \mathcal F_{N_f=3}\lb t\,\bigl|\,c;\{P_k\}\rb=
    \lim_{\Lambda\to\infty} \Lambda^{\Delta_\sigma-\Delta_0-\Delta_t}\mathcal F\lb\substack{\frac{\Lambda+P_*}{2}\;\quad 
    	P_{t}\\\quad  P_\sigma \\ \frac{\Lambda-P_*}{2}\quad P_0};\tfrac{t}{\Lambda}\rb.
    \eeq
    
    The exponentiation conjecture for the quasiclassical asymptotics of the confluent CB \eqref{confCBNf3} is similar to the regular case but involves in addition a rescaling of $t$. More specifically,
    \beq\label{quasiclNf3}
    \mathcal F_{N_f=3}\lb \tfrac{it}{b}\,\bigl|\,c;\tfrac{i\theta_0}{b},\tfrac{i\theta_t}{b},\tfrac{i\theta_*}{b},
    \tfrac{i\sigma}{b}\rb\;\stackrel{b\to 0}{\sim}\; \operatorname{const}\cdot \exp\left\{{b^{-2}\mathcal W_{N_f=3}\lb t\,\bigl|\,\theta_0,\theta_t,\theta_*,\sigma\rb}\right\},
    \eeq
    where the constant (independent of $t$) prefactor does not necessarily have a finite limit as $b\to0$.
    Using \eqref{clCBNf3}, one finds 
    \beq
    \label{NSpNf3}
    \begin{aligned}
    &\,\mathcal W_{N_f=3}\lb t\,\bigl|\,\theta_0,\theta_t,\theta_*,\sigma\rb=\lb\delta_\sigma-\delta_0-\delta_t\rb\ln t-\frac{\lb\delta_\sigma-\delta_0+\delta_t\rb\theta_*}{2\delta_\sigma}\,t+\\
    &+\left[\frac{\lb\delta_\sigma^2-\lb \delta_0-\delta_t\rb^2\rb \theta_*^2}{16\delta_\sigma^3}-\frac{\lb 3\theta_*^2+\delta_\sigma\rb\lb \delta_\sigma^2+2\delta_\sigma\lb\delta_0+\delta_t\rb-3\lb\delta_0-\delta_t\rb^2\rb}{16\delta_\sigma^2\lb 3+4\delta_\sigma\rb}\right]t^2+O\lb t^3\rb,
    \end{aligned}
    \eeq
    where $\delta_0=\frac14-\theta_0^2$, $\delta_t=\frac14-\theta_0^2$, $\delta_\sigma=\frac14-\sigma^2$  as above.
    
     \subsubsection{$N_f=0,1,2$} 
    Let $\Pi_{\Delta}$ denote the identity operator on $\mathcal V_{\Delta}$, considered as an element of $\mathcal V_{\Delta}\otimes \mathcal V_{\Delta}^*$. For instance, writing the contributions of descendant states up to level 2, we have
    \beq\label{resid}
    \Pi_\Delta=|\Delta\rangle \langle \Delta|+\frac{ L_{-1}|\Delta\rangle \langle \Delta|L_1}{2\Delta}+   
    \lb\;L_{-1}^2|\Delta\rangle\;\;L_{-2}|\Delta\rangle\;\rb
    \lb\begin{array}{cc}
    4\Delta\lb 2\Delta+1\rb  & 6\Delta \\ 6\Delta & 4\Delta+\frac{c}{2}
    \end{array}\rb^{-1}\lb\begin{array}{c}
    \langle \Delta|L_1^2 \\ \langle \Delta|L_2
\end{array}\rb+\ldots
    \eeq 
     Confluent CBs of the 1st kind corresponding to $N_f=0,1,2$ are defined as matrix elements
    \begin{subequations}
    	\label{confCBs20}
   	  \begin{align}
   	  \label{confCBNf2}
   	\mathcal F_{N_f=2}\lb t\,\bigl|\,c;P_\star,P_*,P_\sigma\rb:=&\,\left\langle P_*,\tfrac14\bigl|t^{L_0}\Pi_{\Delta_\sigma}\bigr|P_\star,\tfrac14\right\rangle,\\
   	\label{confCBNf1}
   	\mathcal F_{N_f=1}\lb t\,\bigl|\,c;P_*,P_\sigma\rb:=&\,\left\langle P_*,\tfrac14\bigl|t^{L_0}\Pi_{\Delta_\sigma}\bigr|1,0\right\rangle,\\
   		\label{confCBNf0}
   	\mathcal F_{N_f=0}\lb t\,\bigl|\,c;P_\sigma\rb:=&\,\left\langle 1,0\bigl|t^{L_0}\Pi_{\Delta_\sigma}\bigr|1,0\right\rangle.
   	\end{align}
   	\end{subequations}
    Small $t$ expansions of these CBs can be computed using \eqref{resid}:
    \begin{subequations}
    	\label{confCBexps}
   	  	  \begin{align}
   	\label{confCBNf2exp}
   	\mathcal F_{N_f=2}\lb t\,\bigl|\,c;P_\star,P_*,P_\sigma\rb=&\,t^{\Delta_\sigma}\left[1+\frac{P_*P_\star}{2\Delta_\sigma}t+\frac{P_\star^2P_*^2\lb \frac{c}{4\Delta_\sigma}+2\rb-\frac34\lb P_\star^2+P_*^2\rb+\frac18\lb 1+2\Delta_\sigma\rb}{16\Delta_\sigma^2-10\Delta_\sigma+c\left(1+2\Delta_\sigma\right)} \, t^2+O\lb t^3\rb\right],\\
   	\label{confCBNf1exp}
   	\mathcal F_{N_f=1}\lb t\,\bigl|\,c;P_*,P_\sigma\rb=&\,t^{\Delta_\sigma}\left[1+\frac{P_*}{2\Delta_\sigma}t+
   	\frac{P_*^2\lb \frac{c}{4\Delta_\sigma}+2\rb-\frac34}{16\Delta_\sigma^2-10\Delta_\sigma+c\left(1+2\Delta_\sigma\right)} \, t^2+O\lb t^3\rb\right],\\
   	\label{confCBNf0exp}
   	\mathcal F_{N_f=0}\lb t\,\bigl|\,c;P_\sigma\rb=&\,t^{\Delta_\sigma}\left[1+\frac{t}{2\Delta_\sigma}+\frac{ \frac{c\;}{4\Delta_\sigma}+2}{16\Delta_\sigma^2-10\Delta_\sigma+c\left(1+2\Delta_\sigma\right)} \, t^2+O\lb t^3\rb\right].
   	\end{align}
   	\end{subequations}
   	The same expansions are also generated by a chain of termwise confluent  limits:
   	\begin{subequations}
   		\label{cfchain}
   	\begin{align}
   	\mathcal F_{N_f=2}\lb t\,\bigl|\,c;P_\star,P_*,P_\sigma\rb=&\,\lim_{\Lambda\to\infty}
   	\Lambda^{\Delta_\sigma-\Delta_0-\Delta_t}t^{\Delta_0+\Delta_t}\mathcal F_{N_f=3}\lb \tfrac{t}{\Lambda}\,\bigl|\,c;P_0=\tfrac{\Lambda-P_\star}{2},P_t=\tfrac{\Lambda+P_\star}{2},P_*,P_\sigma\rb,\\
   	\mathcal F_{N_f=1}\lb t\,\bigl|\,c;P_*,P_\sigma\rb=&\,\lim_{P_\star\to\infty} 
   	P_\star^{\Delta_\sigma}\mathcal F_{N_f=2}\lb \tfrac{t}{P_\star}\,\bigl|\,c;P_\star,P_*,P_\sigma\rb,\\
   	\mathcal F_{N_f=0}\lb t\,\bigl|\,c;P_\sigma\rb=&\,\lim_{P_*\to\infty}P_*^{\Delta_\sigma}\mathcal F_{N_f=1}\lb \tfrac{t}{P_*}\,\bigl|\,c;P_*,P_\sigma\rb.
   	\end{align}
   	\end{subequations}
   
   The exponentiation  of the confluent CBs \eqref{confCBs20} is described by conjectural formulas analogous to \eqref{quasiclNf3}:
   \begin{subequations}
     \begin{align}
     \label{quasiclNf2}
  \mathcal F_{N_f=2}\lb -\tfrac{t}{b^2}\,\bigl|\,c;\tfrac{i\theta_\star}{b},\tfrac{i\theta_*}{b},
  \tfrac{i\sigma}{b}\rb\;\stackrel{b\to 0}{\sim}&\; \operatorname{const}\cdot \exp\left\{{b^{-2}\mathcal W_{N_f=2}\lb t\,\bigl|\,\theta_\star,\theta_*,\sigma\rb}\right\},\\
       \label{quasiclNf1}
  \mathcal F_{N_f=1}\lb -\tfrac{it}{b^3}\,\bigl|\,c;\tfrac{i\theta_*}{b},
  \tfrac{i\sigma}{b}\rb\;\stackrel{b\to 0}{\sim}&\; \operatorname{const}\cdot \exp\left\{{b^{-2}\mathcal W_{N_f=1}\lb t\,\bigl|\,\theta_*,\sigma\rb}\right\},\\
   \label{quasiclNf1}
  \mathcal F_{N_f=0}\lb \tfrac{t}{b^4}\,\bigl|\,c;\tfrac{i\sigma}{b}\rb\;\stackrel{b\to 0}{\sim}&\; \operatorname{const}\cdot \exp\left\{{b^{-2}\mathcal W_{N_f=0}\lb t\,\bigl|\,\sigma\rb}\right\}
  \end{align}
   \end{subequations}
   These confluent variants of Conjecture A are of course straightforward to check at low orders in $t$. We record below only a few terms in the expansions of $\mathcal W_{N_f=0,1,2}\lb t\rb$ which easily follow from \eqref{confCBexps},
   \begin{subequations}
   \begin{align}
   \mathcal W_{N_f=2}\lb t\,\bigl|\,\theta_\star,\theta_*,\sigma\rb=&\,\delta_\sigma\ln t+
   \frac{\theta_\star\theta_*}{2\delta_\sigma}\,t+\lb
   \frac{\lb 3\theta_*^2+\delta_\sigma\rb \lb 3\theta_\star^2+\delta_\sigma\rb}{16\delta_\sigma^2\lb 3+4\delta_\sigma\rb}-\frac{\theta_*^2\theta_\star^2}{16\delta_\sigma^3}\rb t^2+O\lb t^3\rb,\\
   \mathcal W_{N_f=1}\lb t\,\bigl|\,\theta_*,\sigma\rb=&\,\delta_\sigma\ln t+
   \frac{\theta_*}{2\delta_\sigma}\,t+\frac{\lb 5\delta_\sigma-3\rb\theta_*^2+3\delta_\sigma^2}{16\delta_\sigma^3\lb 3+4\delta_\sigma\rb} t^2+O\lb t^3\rb,\\
   \mathcal W_{N_f=0}\lb t\,\bigl|\,\sigma\rb=&\,\delta_\sigma\ln t+
   \frac{t}{2\delta_\sigma}+\frac{ 5\delta_\sigma-3}{16\delta_\sigma^3\lb 3+4\delta_\sigma\rb} t^2+O\lb t^3\rb.
   \end{align}
   \end{subequations}
   Note that the expansions of $\mathcal W_{N_f<4}\lb t\rb$ can also be obtained directly  from  the regular quasiclassical CB $\mathcal W\lb t\bigl|\left\{\delta_k \right\}\rb$ by the same sequence of termwise limits as in \eqref{confchain}; in other words, for CBs of the 1st kind the quasiclassical limit commutes with confluence.

   	\subsection{Confluent conformal blocks of the 2nd kind\label{subsec2kind}}
   	\subsubsection{Irregular vertex operators}
    Let us recall the construction of a (dual) irregular vertex operator $V_{\vec\lambda,\vec\lambda'}^{\Delta}\lb t\rb:\mathcal V_{\vec\lambda}^{W,*}\to
    \mathcal V_{\vec\lambda'}^{W,*}$ from \cite{Nagoya1}. It  is uniquely determined by the following properties:
    \begin{itemize}
    	\item Whittaker modules $\mathcal V_{\vec\lambda}^{W,*}$, $\mathcal V_{\vec\lambda'}^{W,*}$ have the same rank $r$ and satisfy a genericity condition $\lambda_{2r},\lambda_{2r}'\ne 0$.
    	\item The operator $V_{\vec\lambda,\vec\lambda'}^{\Delta}\lb t\rb$ has the same commutation relations \eqref{VOVir} with the Virasoro generators as the usual vertex operator:
    	\beq
    	\left[ L_n, V_{\vec\lambda,\vec\lambda'}^{\Delta}\lb t\rb
    	\right]=t^n\left( t \tfrac{\partial}{\partial t}+
    	\lb n+1\rb\Delta\right)
    	V_{\vec\lambda,\vec\lambda'}^{\Delta}\lb t\rb.
    	\eeq
    	This is essentially an expression of the local conformal invariance.
    	\item Most importantly, there is a ``normalization'' condition
    	\beq\label{normirrVO}
    	\bigl\langle \lambda_r,\ldots,\lambda_{2r}\bigr|V_{\vec\lambda,\vec\lambda'}^{\Delta}\lb t\rb=
    	t^{\alpha}\exp\left\{\sum\nolimits_{k=1}^r \beta_k t^k\right\}\left[\bigl\langle \lambda_r',\ldots,\lambda_{2r}'\bigr|+\sum_{n=1}^\infty t^{-n}\,\bigl\langle w'_n\bigr|\,\right],
    	\eeq
    	where the expression in the square brackets is understood as a formal series in  $ \mathcal V_{\vec\lambda'}^{W,*}\otimes\mathbb C[[t^{-1}]]$.
    \end{itemize}
    The parameters $\vec{\lambda}'$, $\alpha$, $\beta_1,\ldots,\beta_{r-1}$ on the right of \eqref{normirrVO} as well as all coefficients $\bigl\langle w'_n\bigr|\in \mathcal V_{\vec\lambda'}^{W,*}$ of the formal series are determined by $\vec\lambda$, $\Delta$ and $\beta_r$. In particular, one has
    \beq
    \lambda'_k=\begin{cases}
    \;\lambda_k,\qquad & k=r+1,\ldots, 2r,\\ \lambda_r-r\beta_r,\qquad & k=r.\end{cases}
    \eeq  
    
    \subsubsection{Confluent conformal blocks of type $\mathcal D$}
    The canonical pairing between the Virasoro highest weight modules and Whittaker modules of rank 1, already used in the previous subsection, allows to define \cite{Nagoya1} a class of 3-point confluent CBs of the 2nd kind:
    \beq
    \mathcal{D}_{N_f=3}\lb t\,\bigl|\,c;P_0,P_t,P_*,P_\nu\rb:=\Bigl\langle P_*,\tfrac14\Bigl|V_{\lb P_*,\frac14\rb,\lb P_\nu,\frac14\rb}^{\Delta_t}\lb t\rb\Bigr|\Delta_0\Bigr\rangle=\vcenter{\hbox{ \includestandalone[scale=1]{conflBlock2}}}
    \eeq
    They will be referred to as confluent CBs of type $\mathcal D$. The coefficients of the large~$t$ expansion of these CBs, 
    \beq
    \mathcal{D}_{N_f=3}\lb t\,\bigl|\,c;\left\{P_k\right\}\rb=t^{2P_\nu\lb P_*-P_\nu\rb}e^{\,\lb P_*-P_\nu\rb t}\left[1+\sum_{n=1}^{\infty}\mathcal D_n\lb c;\left\{P_k\right\}\rb t^{-n}\right],
    \eeq
    as well as the exponent $\alpha=2P_\nu\lb P_*-P_\nu\rb$, are thus fixed by Virasoro symmetry. Two lowest order coefficients can be deduced from the expressions for $\langle w'_1|$, $\langle w'_2|$ in \cite[Appendix A.2.1]{Nagoya1}. One has
    \begin{subequations}
    	\label{coefsD12}
    \begin{align}
    \label{coefD1}
    \mathcal D_1\lb c;\left\{P_k\right\}\rb=&\,-4P_\nu^3+6P_\nu^2P_*+2P_\nu\lb\Delta_0+\Delta_t-P_*^2\rb-2\Delta_0 P_*,\\
    \label{coefD2}
    \mathcal D_2\lb c;\left\{P_k\right\}\rb=&\,\tfrac12\mathcal D_1\lb c;\left\{P_k\right\}\rb^2+2\bigl( \Delta_0+\lb P_*-3P_\nu\rb P_\nu\bigr)\, \bigl( \Delta_t-\lb 2P_*-3P_\nu\rb\lb P_*-P_\nu\rb\bigr)+\\
    \nonumber+&\,\lb 2\lb P_*-2P_\nu\rb^2+\tfrac{c-1}{3}\rb P_\nu\lb P_*-P_\nu\rb.
    \end{align}
    \end{subequations}
    
    The algebraic computation of the coefficients $\left\{\mathcal D_k\right\}$ becomes quite lengthy at higher orders. However, all of them remain explicitly computable (at least in principle) since $\mathcal{D}_{N_f=3}\lb t\,\bigl|\,c;\left\{P_k\right\}\rb$ can be identified \cite{LNR} with a particular limit of the regular $u$-channel CB suggested in \cite{GT}. Namely,
    \beq\label{GTlimit}
    1+\sum_{n=1}^{\infty}\mathcal D_n\lb c;\left\{P_k\right\}\rb t^{-n}=\lim_{\Lambda\to\infty}\lb\frac{\Lambda}{t}\rb^{\Delta_t-\lb P_*-P_\nu\rb\lb \Lambda-P_\nu\rb}\lb 1-\frac{\Lambda}{t}\rb^{\lb P_*-P_\nu\rb\lb\Lambda+P_\nu\rb+\Delta_t} \mathcal F\lb\substack{\frac{\Lambda+P_*}{2}\qquad\quad\; \;
    	P_{t}\;\;\\ \quad  \frac{\Lambda+P_*}{2}-P_\nu \\ \quad P_0 \quad\qquad\;\; \frac{\Lambda-P_*}{2}};\tfrac{\Lambda}{t}\rb,
    \eeq
    where the last two factors under the limit are interpreted as formal series in $\frac1t$ (note that only the \textit{product} of these series admits a finite limit as $\Lambda\to\infty$ while the coefficients of each of them diverge when treated separately). Another difference from the confluent limit leading to CBs of the 1st kind is that the intermediate dimension also goes to $\infty$. The relation \eqref{GTlimit} of course reproduces \eqref{coefD1}, \eqref{coefD2} and allows to go to any desired order. For instance,
    \begin{align}
    &\mathcal D_3\lb c;\left\{P_k\right\}\rb=\mathcal D_1\lb c;\left\{P_k\right\}\rb\mathcal D_2\lb c;\left\{P_k\right\}\rb-\tfrac13 \mathcal D_1\lb c;\left\{P_k\right\}\rb^3-\Bigl( 11P_\nu\lb P_*-P_\nu\rb-\tfrac{5}{3}P_*^2+\tfrac{13}{6}\lb\Delta_0+\Delta_t\rb\Bigr)\, \mathcal D_1\lb c;\left\{P_k\right\}\rb+\\
    \nonumber
    &\;+\tfrac{11}{3}\Bigl(\Delta_t-\Delta_0-2P_*\lb P_*-2P_\nu\rb\Bigr)\Bigl(\lb P_*-P_\nu\rb \Delta_0+P_\nu\Delta_t+P_*P_\nu\lb P_*-P_\nu\rb\Bigr)+\tfrac23\lb c-2\rb\Bigl(\lb P_*-P_\nu\rb \Delta_0-P_\nu\Delta_t\Bigr)+\\
    \nonumber &\;+\tfrac19\lb 66P_*^2+17c-23\rb\lb P_*-2P_\nu\rb\lb P_*-P_\nu\rb P_\nu.
    \end{align}
    
    The above expressions for $\mathcal D_{2,3}$ are organized in a way which makes manifest the exponentiation of CBs of type $\mathcal D$ at low orders. We thus arrive at the following confluent variant of Conjecture A, cf \eqref{quasiclNf3}.
    \begin{conj}
    Confluent conformal blocks  of type $\mathcal D$ exponentiate as
     \beq\label{quasiclNf3bis}
    \mathcal D_{N_f=3}\lb \tfrac{it}{b}\,\bigl|\,c;\tfrac{i\theta_0}{b},\tfrac{i\theta_t}{b},\tfrac{i\theta_*}{b},
    \tfrac{i\nu}{b}\rb\;\stackrel{b\to 0}{\sim}\; \operatorname{const}\cdot \exp\left\{{b^{-2}\mathcal U_{N_f=3}\lb t\,\bigl|\,\theta_0,\theta_t,\theta_*,\nu\rb}\right\},
    \eeq
    with
    \beq
    \mathcal U_{N_f=3}\lb t\,\bigl|\,\theta_0,\theta_t,\theta_*,\nu\rb=\lb\nu-\theta_*\rb t+2\nu\lb\nu-\theta_*\rb\ln t+\sum_{n=1}^\infty 
    \mathcal U_n\lb \theta_0,\theta_t,\theta_*,\nu\rb t^{-n}.
    \eeq
    \end{conj}
   We have checked this claim up to order $O\lb t^{-4}\rb$. The coefficients of the classical CB  of type $\mathcal D$ are readily obtained from those of $\mathcal{D}_{N_f=3}\lb t\rb$. E.g. from \eqref{coefsD12} it follows that
   \begin{subequations}
   	\label{U12}
   	\begin{align}
   	\mathcal U_1\lb \theta_0,\theta_t,\theta_*,\nu\rb=&\,4\nu^3-6\nu^2\theta_*+2\nu\lb\delta_0+\delta_t+\theta_*^2\rb   -2\delta_0\theta_*,\\
   	\mathcal U_2\lb\theta_0,\theta_t,\theta_*,\nu\rb=&\,-2\bigl( \delta_0-\nu\lb\theta_*-3\nu\rb\bigr)\bigl(\delta_t+\lb 2\theta_*-3\nu\rb\lb\theta_*-\nu\rb\bigr)-2\nu\lb\theta_*-\nu\rb\lb \lb\theta_*-2\nu\rb^2-1\rb.	\end{align}
   \end{subequations}
    Assuming that the quasiclassical limit commutes with confluence for the CBs of the 2nd kind as well, we can alternatively compute these coefficients from a quasiclassical version of \eqref{GTlimit},
    \beq 
    \mathcal U_n\lb \theta_0,\theta_t,\theta_*,\nu\rb =\lim_{\Lambda\to\infty}\Lambda^n\left[
    \mathcal W_n\lb\substack{\frac{\Lambda+\theta_*}{2}\qquad\quad
    	\theta_{t}\;\;\\ \;\;  \frac{\Lambda+\theta_*}{2}-\nu \\ \quad \theta_0 \quad\qquad \frac{\Lambda-\theta_*}{2}}\rb-\frac{\delta_t-\lb\theta_*-\nu\rb\lb\Lambda+\nu\rb}{n} \right],
    \eeq
    where $\mathcal W_n$'s denote the expansion coefficients of the regular classical CB.

    \subsubsection{Confluent conformal blocks of type $\mathcal G$}
    There is no obvious canonical pairing between Whittaker modules of rank $r\ge2$ and highest weight representations. For example, a pairing such as $\bigl\langle\vec\lambda\bigr|\cdot L_{-1}^{k}\bigl|\Delta\bigr\rangle$ of the Whittaker vector and a descendant state in $\mathcal V_\Delta$ for $r>1$ cannot be reduced to ${\bigl\langle\vec\lambda\bigr|\Delta\bigl\rangle}$ since $\bigl\langle\vec{\lambda}\bigr|$ is not an eigenstate of $L_{-1}$. A notable exception concerns the pairing of Whittaker modules of rank $2$ with an \textit{irreducible} highest weight module $\mathcal V_0$ (as in this case $L_{-1}|0\rangle=0$). This allows to define \cite{Nagoya1} confluent CBs 
    \beq
    \label{CBG}
    \mathcal G\lb t\,|\,c;P_0,P_\bullet,P_\nu\rb=\bigl\langle P_\bullet,0,\tfrac14\bigr|
    V^{\Delta_0}_{\lb P_\bullet,0,\tfrac14\rb,\lb P_\bullet-P_\nu,0,\tfrac14\rb}\lb t\rb\bigl|0\bigr\rangle=\vcenter{\hbox{\includestandalone[height=2.2cm]{conflBlock3}}}
    \eeq
    They will be referred to as CBs of the 2nd kind of type $\mathcal G$. Setting the eigenvalues $\lambda_{3}$, $\lambda_4$ to $0$ and $\tfrac14$ involves no loss of generality thanks to translation and dilatation symmetry of \eqref{CBG}. 
    
    The large $t$ expansion of the conformal block $\mathcal G\lb t|\,c;\left\{P_k\right\}\rb$ has the form
    \beq
    \mathcal G\lb t|\,c;\left\{P_k\right\}\rb=
    t^{\Delta_0-3P_\nu^2+2P_\bullet P_\nu}e^{\frac{P_\nu t^2}2}
    \left[1+\sum_{n=1}^{\infty}
    \mathcal G_n\lb c;\left\{P_k\right\}\rb t^{-2n}\right].
    \eeq
    Algebraic definition \eqref{CBG} determines the exponent $\tilde\alpha=\Delta_0-3P_\nu^2+2P_\bullet P_\nu$ in the above and also allows to compute the coefficients $\mathcal G_n$. One has, in particular,
    \begin{subequations}
    \begin{align}
    \mathcal G_1\lb c;\left\{P_k\right\}\rb=&\,6P_\nu^3-6P_\bullet P_\nu^2+\lb P_\bullet^2-3\Delta_0-\tfrac{c-1}{8}\rb P_\nu+2P_\bullet\Delta_0,\\
    \mathcal G_2\lb c;\left\{P_k\right\}\rb=&\,\tfrac12 \mathcal G_1\lb c;\left\{P_k\right\}\rb^2+\tfrac{105}{4}P_\nu^4-35P_\bullet P_\nu^3+\lb 12 P_\bullet^2-\tfrac{33}{2}\Delta_0-\tfrac{13c-19}{8}\rb P_\nu^2+\\
    \nonumber
    +&\,\lb-P_\bullet^2+18\Delta_0+\tfrac{19c-31}{24}\rb P_\bullet P_\nu+\tfrac14\lb-16P_\bullet^2+\Delta_0+c-2\rb\Delta_0.
    \end{align}
    \end{subequations}
    \begin{conj}
    	Confluent conformal blocks of type $\mathcal G$ exponentiate as
       \beq\label{quasiclCBG}
    \mathcal G\lb t\sqrt{\tfrac{2i}{b}}\,\bigl|\,c;\tfrac{i\theta_0}{b},\tfrac{i\theta_\bullet}{b},
    \tfrac{i\nu}{b}\rb\;\stackrel{b\to 0}{\sim}\; \operatorname{const}\cdot \exp\left\{b^{-2}\tilde{\mathcal U}\lb t\,\bigl|\,\theta_0,\theta_\bullet,\nu\rb\right\},
    \eeq
    with
    \beq\label{Utccc}
    \tilde{\mathcal U}\lb t\,\bigl|\,\theta_0,\theta_\bullet,\nu\rb=-{\nu t^2} +\lb \delta_0+3\nu^2-2\theta_\bullet\nu\rb\ln t+\sum_{n=1}^\infty 
    \tilde{\mathcal U}_n\lb \theta_0,\theta_\bullet,\nu\rb t^{-2n}.
    \eeq
    \end{conj}
   \noindent Let record the first coefficients of the classical CB of type $\mathcal G$:
    \begin{subequations}
    	\label{Ut12}
    \begin{align}
    \tilde{\mathcal U}_1\lb \theta_0,\theta_\bullet,\nu\rb=&\,-\tfrac12\left[6\nu^3-6\theta_\bullet\nu^2+\lb\theta_\bullet^2+3\delta_0+\tfrac34\rb\nu-2\delta_0\theta_\bullet\right],\\
 \tilde{\mathcal U}_2\lb \theta_0,\theta_\bullet,\nu\rb=&\,-\tfrac14\left[\tfrac{105}{4}\nu^4-35\theta_\bullet\nu^3 +\lb12\theta_\bullet^2+\tfrac{33}{2}\delta_0+\tfrac{39}{4}\rb\nu^2-\lb\theta_\bullet^2+18\delta_0+\tfrac{19}{4}\rb\theta_\bullet \nu  +\lb 4 \theta_\bullet^2+\tfrac14\delta_0+\tfrac32\rb \delta_0\right].
    \end{align}
    \end{subequations}
    In the next section, they will be compared with the coefficients of the 
    strong-coupling expansion of accessory parameter function of the biconfluent Heun equation.

    \section{Accessory parameter functions of Bohr-Sommerfeld type}
    \subsection{Bohr-Sommerfeld periods}
    Consider the Schroedinger equation 
    \beq
    \hbar^2\psi''\lb z\rb=U\lb z\rb \psi\lb z\rb,
    \eeq 
    with a rational potential $U\lb z\rb$ which can also analytically depend on $\hbar$, so that $U\lb z\rb=\sum_{n=0}^\infty \hbar^n U_n\lb z\rb$ with $U_0\lb z\rb\ne 0$. It admits a formal WKB solution
    \beq
    \psi\lb z\rb=\exp\left\{\sum_{n=-1}^\infty \hbar^{n} \int^z S_n\lb z\rb dz \right\},
    \eeq  
    where $S_{-1}\lb z\rb\equiv p\lb z\rb=\ds\sqrt{\ds U\lb z\rb}$  and all other
    $S_{k}$'s are determined by the recurrence relations
    \beq
    S_{n}'\lb z\rb+\sum_{k=-1}^{n+1}S_k\lb z\rb S_{n-k}\lb z\rb=0,\qquad n \ge -1.
    \eeq
    In particular, one has
    \begin{subequations}
    \begin{alignat}{2}
     & S_0=\lb-\tfrac12\ln p\rb', && S_1=\frac{2pp''-3\lb p'\rb^2}{8p^3}\\
     & S_2=\,\lb\frac{3\lb p'\rb^2}{16 p^4}-\frac{p''}{8p^3}\rb',\qquad && 
     S_3=-\frac{297\lb p'\rb^4-396 p\lb p'\rb^2 p''+52p^2\lb p''\rb^2+80p^2p'p'''-8p^3p''''}{128p^7},
    \end{alignat}
    \end{subequations}
    and so on. All even contributions $S_{2n}\lb z\rb$  are  given by total derivatives. In fact, denoting
    \beq
    S_{\mathrm{odd}}\lb z\rb:=\sum_{n=0}^\infty \hbar^{2n-1}S_{2n-1}\lb z\rb,\qquad S_{\mathrm{even}}\lb z\rb:=\sum_{n=0}^\infty \hbar^{2n}S_{2n}\lb z\rb, 
    \eeq
    it can be shown that $S'_{\mathrm{odd}}+2S_{\mathrm{even}}S_{\mathrm{odd}}=0$, and therefore $\psi\lb z\rb=\frac{1}{\sqrt{S_{\mathrm{odd}}\lb z\rb}}\exp\int^zS_{\mathrm{odd}}\lb z\rb dz$. 
      
    Given a cycle~$\gamma$ on the Riemann surface $p^2=U\lb z\rb$, one may then introduce the \textit{Bohr-Sommerfeld (BS) period}, describing a formal monodromy of $\psi\lb z\rb$ along $\gamma$:
    \beq
    \nu:=\frac{1}{2\pi i}\oint_\gamma S_{\mathrm{odd}}\lb z\rb dz= \sum_{n=0}^\infty\hbar^{2n-1}\nu_{2n-1},
    \eeq
    where we denote
    $\nu_{2n-1}:=\ds\frac{1}{2\pi i}\ds\oint_\gamma S_{2n-1}\lb z\rb dz$.

   \subsection{Confluent Heun} 
    Let us now consider the potential of H$_{\mathrm{V}}$,
    \beq 
    V\lb z\rb=-\frac{\delta_0}{z^2}-\frac{\delta_t}{\lb z-t\rb^2}+\frac14+\frac{\theta_*}{z}-\frac{\mathcal E}{z\lb z-t\rb}=\frac{P_4\lb z\rb}{z^2\lb z-t\rb^2}
    \eeq
    and explain how a small parameter playing the role of $\hbar$ appears in our setup. Suppose that $t$ is large and that $\mathcal E/t=\kappa +o\lb 1\rb$. It is not difficult to understand that in this case two roots $z_{1,2}$ of the 4th degree polynomial $P_4\lb z\rb$ are of order $O\lb 1\rb$ and two other roots $z_{3,4}$ are close to $t$. The former two turning points are asymptotically close to the roots of $-\frac{\delta_0}{z^2}+\frac14+\frac{\theta_*+\kappa}{z}=0$, the latter are asymptotic to $t+\xi$, where $\xi$ satisfies $-\frac{\delta_t}{\xi^2}+\frac14-\frac{\kappa}{\xi}=0$. To make the differentials $S_{n}\lb z\rb dz$ single-valued on $\mathbb C\mathbb P^1$, we introduce two branch cuts: one connecting $z_1$ and $z_2$ and another connecting $z_3$ and $z_4$.
    
    Choose the cycle $\gamma$ to be the circle of radius $\sqrt{t}$, so that the branch points $z=z_{1,2}$ (and eventually the pole $z=0$) are inside $\gamma$, while the branch points  $z=z_{3,4}$ (and eventually the poles $z=t,\infty$) are outside. Equivalently, we could set from the very beginning $z= \lambda\sqrt t$, so that in the $\lambda$-plane the first two branch points would become $\lambda_{1,2}=O\lb\frac{1}{\sqrt t}\rb$ and two others $\lambda_{3,4}=O\lb \sqrt t\rb$. The cycle $\gamma$ maps to the unit circle $|\lambda|=1$ and the potential becomes
    $tV\lb \lambda\sqrt t \rb=\hbar^{-2}U\lb\lambda\rb$, with $\hbar=\frac{1}{\sqrt t}$ and
    \beq
     U\lb \lambda\rb=\frac14+\frac{\hbar\theta_*}{\lambda}+\frac{\hbar^3\mathcal E}{\lambda\lb 1-\hbar\lambda\rb}-\frac{\hbar^2\delta_0}{\lambda^2}-\frac{\hbar^4\delta_t}{\lb 1-\hbar\lambda\rb^2}.
    \eeq 
    Recall that $\hbar^2\mathcal E=\kappa+o\lb 1\rb$, and therefore $U\lb \lambda\rb=\frac14+O\lb\hbar\rb$. In order to calculate different contributions to the BS period $\nu$, it now suffices to expand the integrands in $\hbar$ and compute the residue  at $0$ or $\infty$. Substituting
    \beq\label{EnAns}
    \mathcal E= \kappa t+\sum_{n=0}^{\infty}\mathcal E_n t^{-n}
 ,
    \eeq
    we find that
    \begin{subequations}
    \begin{align}
    &\begin{aligned}
    \nu_{-1}=&\,\frac1{2\pi i}\oint_{|\lambda|=1}p\lb\lambda\rb\,d\lambda=
    \lb \kappa+\theta_*\rb\hbar +\lb \mathcal E_0-2\kappa^2-2\kappa\theta_*\rb \hbar^3+\\
    &\,+\Bigr[\mathcal E_1-2\mathcal E_0\lb\theta_*+2\kappa\rb+2\kappa\lb 6\kappa^2+3\theta_*^2+9\kappa\theta_*+\delta_0+\delta_t\rb+2\delta_t\theta_*\Bigl]\hbar^5 +\\
    &\,+\Bigr[\mathcal E_2-2\mathcal E_1\lb\theta_*+2\kappa\rb+2\mathcal E_0\lb-\mathcal E_0+\delta_0+\delta_t+3\theta_*^2+18\kappa\theta_*+18\kappa^2\rb-100\kappa^2\lb\theta_* +\kappa\rb^2\\
    &\quad -12\lb\theta_*+2\kappa\rb\lb \kappa\delta_0+\kappa\delta_t+\delta_t\theta_* \rb -20\theta_*^2\kappa\lb\theta_*+\kappa\rb-4\delta_0\delta_t\Bigl]\hbar^7+ O\lb\hbar^9\rb,
    \end{aligned} \\
    &\nu_{1}\;= \frac1{2\pi i}\oint_{|\lambda|=1} \frac{2pp''-3\lb p'\rb^2}{8p^3}d\lambda=
    -4\kappa\lb \kappa +\theta_*\rb\hbar^5 +O\lb\hbar^7\rb,\\
    &\nu_{3}\,\,=-48\kappa\lb \kappa +\theta_*\rb\hbar^7 +O\lb\hbar^9\rb,\qquad \ldots
    \end{align}
    \end{subequations}
    Note in particular that the leading contribution to the period is of 0th order: $\nu=\kappa+\theta_*+O\lb \hbar\rb$.
    
    Inverting the relation $\nu=\nu\lb\mathcal E,t\rb$ to express $\mathcal E=\mathcal E^{\mathrm{[BS]}}\lb t\,\ds|\,\nu\rb$ as a function of $\nu$ and $t$ is equivalent to asking all quantum ($\hbar$-dependent) corrections to the BS period to vanish. It follows that
    \begin{gather}
    \mathcal E_0=2\kappa\lb \kappa+ \theta_*\rb=2\lb \nu-\theta_*\rb\nu,\qquad
    \mathcal E_1=-\mathcal U_1\lb\theta_0,\theta_t,\theta_*,\nu\rb,\qquad
    \mathcal E_2=-2\mathcal U_2\lb\theta_0,\theta_t,\theta_*,\nu\rb,\quad
    \ldots,
    \end{gather}   	
   	where $\mathcal U_{1,2}$ coincide with those given in \eqref{U12}. It now becomes straightforward to formulate Conjecture B for confluent CBs of type $\mathcal D$.
   	\begin{conj}\label{conjHVCFT}
   		Let $\nu$ be the Bohr-Sommerfeld period along the cycle described above. Then 
   		\beq
   		\mathcal E^{[\mathrm{BS}]}\lb t\,|\,\theta_0,\theta_t,\theta_*,\nu\rb=t\frac{\partial}{\partial t}\mathcal U_{N_f=3}\lb t\,\bigl|\,\theta_0,\theta_t,\theta_*,\nu\rb,
   		\eeq
    where $\mathcal U_{N_f=3}\lb t\,\bigl|\,\theta_0,\theta_t,\theta_*,\nu\rb$ is the classical confluent conformal block of type $\mathcal D$.
   	\end{conj}
   	
   	\subsection{Biconfluent Heun}
   	Let us now follow a similar approach for the biconfluent HE. The potential of H$_{\mathrm{IV}}$ is given by
   	\beq
   	V\lb z\rb= -\frac{\delta_0}{z^2}-\frac{\mathcal E}{z}+ 2\theta_{\bullet}+\lb z+t\rb^2=\frac{Q_4\lb z\rb}{z^2}.
   	\eeq
   	Assume again that $t$ is large and that $\mathcal E/t=\kappa +o\lb 1\rb$. Two zeros $z_{1,2}$ of $Q_4\lb z\rb$ are asymptotic to $\frac{\xi_{1,2}}{t}$, where $\xi_{1,2}$ satisfy the equation $-\frac{\delta_0}{\xi^2}-\frac{\kappa}{\xi}+1=0$. Two other zeros $z_{3,4}$ are asymptotic to $-t+\xi_{3,4}$, where $\xi_{3,4}$ satisfy the equation
   	$\kappa+2\theta_\bullet+\xi^2=0$. Introducing branch cuts going from $z_1$ to $z_2$ and from $z_3$ to $z_4$, we choose  $\gamma$ defining the BS period to be the unit circle. Also, we rewrite the potential as $V\lb z\rb=\hbar^{-2} U\lb z\rb$, with $\hbar=t^{-1}$ and
   	\beq
   	U\lb z\rb= 1+2\hbar z-\frac{\hbar^2\mathcal E}{z}+\hbar^2 \lb z^2-\frac{\delta_0}{ z^2}+2\theta_\bullet\rb.
   	\eeq
   	By the above assumption, $\hbar\mathcal E=\kappa+o\lb 1\rb$. Making the ansatz \eqref{EnAns}, one obtains
   	\begin{subequations}
   	\begin{align}
   	&\begin{aligned}
   	\nu_{-1}=&\,-\tfrac12\,\kappa\hbar-\tfrac12\,\mathcal E_0\hbar^2+\tfrac12\lb-\mathcal E_1+\delta_0+\kappa\theta_\bullet+\tfrac34\,\kappa^2\rb\hbar^3+\tfrac14\lb-2\mathcal E_2+ \lb 2\theta_\bullet+3\kappa\rb\mathcal E_0\rb\hbar^4+\\
   	&\,+\tfrac14\,\Bigl(-2\mathcal E_3+\mathcal E_1\lb 2\theta_\bullet+3\kappa\rb+\tfrac32\mathcal E_0^2-\tfrac{15}{4}\kappa^2\lb 2\theta_\bullet+\kappa\rb-6\delta_0\lb \theta_\bullet+\kappa\rb-3\theta_\bullet^2\kappa\Bigr)\,\hbar^5+O\lb \hbar^6\rb,
   	\end{aligned}\\
   	&\nu_1\;=-\tfrac{3}{16}\kappa\hbar^3-\tfrac{3}{16}\mathcal E_0\hbar^4+\tfrac1{64}\lb-12\mathcal E_1+60\delta_0+100\theta_\bullet\kappa+105\kappa^2\rb\hbar^5+O\lb \hbar^6\rb,\\
   	&\nu_3\,\,=-\tfrac{315}{256}\kappa\hbar^5+O\lb \hbar^6\rb,\quad \ldots
   	\end{align}
   	\end{subequations}
   	Requiring the quantum corrections to the BS period to vanish, we find that
   	\begin{subequations}
   	\begin{gather}
   	\mathcal E_0=\mathcal E_2=\mathcal E_4=\ldots=0,\qquad \kappa=-2 \nu,\qquad
   	\mathcal E_1=\delta_0+3\nu^2-2\theta_\bullet \nu,\\
   	\mathcal E_3=6\nu^3-6\theta_\bullet \nu^2+\lb \theta_\bullet^2+ 3\delta_0+\tfrac34\rb\nu-2\delta_0\theta_\bullet,\qquad \ldots
   	\end{gather}
   	\end{subequations}
   	Compraison of these expressions with \eqref{Utccc}--\eqref{Ut12} finally leads to a variant of Conjecture B for confluent CBs of type $\mathcal G$.
   	\begin{conj}\label{conjHIVCFT}
   		Let $\nu$ be the Bohr-Sommerfeld period along the cycle described above. Then
   			\beq
   	\mathcal E^{[\mathrm{BS}]}\lb t\,|\,\theta_0,\theta_\bullet,\nu\rb=\frac{\partial}{\partial t}\tilde{\mathcal U}\lb t\,\bigl|\,\theta_0,\theta_\bullet,\nu\rb,
   	\eeq	
   	where $\tilde{\mathcal U}\lb t\,\bigl|\,\theta_0,\theta_\bullet,\nu\rb$ denotes the classical confluent conformal block of type $\mathcal G$.
   	\end{conj}
   	
 	\section{Discussion}
 	We conclude by mentioning a few questions that have been left outside the scope of this note.
 	\begin{itemize}
 		\item All-order Bohr-Sommerfeld relation used to define the accessory parameter functions is an analog of the Hill determinant evaluation of the Mathieu characteristic exponent. Although  $\mathcal E^{\text{[BS]}}\lb t\rb$ was so far defined only as a formal asymptotic series, it can be promoted to an actual function of $t$ by relating the resummed BS period to genuine monodromy/Stokes data with the help of exact WKB techniques, see e.g. \cite{IN,AJJRT}. Classical irregular conformal blocks may then be interpreted as generating functions of canonical transformations between $\lb\mathcal E,t\rb$ and a pair of suitable coordinates on confluent Heun monodromy manifolds.
 	 \item It would be interesting to identify conformal blocks whose quasiclassical limit describes accessory parameters of the confluent Heun equations other than H$_\mathrm{V}$ and H$_\mathrm{IV}$. A recent work \cite{Nagoya2} contains a few promising candidates for this role for H$_{\mathrm{III}_1}$ and H$_{\mathrm{II}}$ (see also \cite{NU}). Even in the case of H$_\mathrm{V}$ and H$_\mathrm{IV}$, there exist other possibilities to define BS accessory parameter function which lead to ans\"atze different from \eqref{EnAns}. On the gauge theory side, they correspond to different strongly coupled expansion points. An algebraic construction of the corresponding CBs of the 2nd kind does not seem to be known.
 	 \item There is a well-known correspondence between Heun and Painlevé equations under which Heun couplings~$t$ are mapped to special points of Painlevé functions. It was realized in \cite{NC,CCN,LN} that this relation can be used to compute asymptotic expansions of Heun accessory parameters. In the terminology of the present paper, \cite{NC,CCN,LN} deal with the weak coupling expansions of Floquet characteristics of H$_{\mathrm{VI}}$ and H$_{\mathrm{V}}$ which can in fact be derived more easily as explained in Appendices~\ref{appendixHeun} and~\ref{appCH}. However the method of \textit{loc. cit.} can be extended to strong coupling using the long-distance Painlevé asymptotic expansions from \cite{BLMST}; this was effectively done for Mathieu equation H$_{\mathrm{III}_3}$ in \cite{GMS}. The Heun/Painlevé correspondence in the latter case admits an interpretation in terms of the quasiclassical limit of the Nakajima-Yoshioka blowup equations for pure $\mathrm{SU}\lb 2\rb$ gauge theory \cite{GG}, whose generalizations to the regular case were recently  studied in \cite{JN,Nekrasov2}. It should be possible to obtain (and perhaps even prove!) similar identities (implicitly present in the approach of \cite{NC,CCN,LN}) relating BS accessory parameters and Painlevé functions in other cases, including those corresponding to Argyres-Douglas theories. We plan to return to these questions in a future work.
 	\end{itemize}
  	
  	\noindent{\small \textbf{Acknowledgements}. The authors are grateful to Pavlo Gavrylenko and Nikolai Iorgov for illuminating discussions, and to Hajime Nagoya for sharing a Mathematica code computing higher order contributions to confluent conformal blocks of type  $\mathcal G$.}
   	
   	\appendix
   	\section{Heun accessory parameter from continued fractions\label{appendixHeun}}
   	It will be convenient for us here to transform the Heun equation \eqref{heune} into its \textit{canonical form} by the substitution
   	\beq
   	\psi\lb z\rb=z^{\frac12-\theta_0}\lb z-t\rb^{\frac12-\theta_t}\lb z-1\rb^{\frac12-\theta_1}\phi\lb z\rb.
   	\eeq
   	The resulting equation for $\phi\lb z\rb$ is given by
  \beq\label{heunecan}
  \phi''\lb z\rb+\lb\frac{\gamma}{z}+\frac{\delta}{z-1}+\frac{\epsilon}{z-t}\rb\phi'\lb z\rb+\frac{\alpha\beta z-q}{z\lb z-1\rb\lb z-t\rb}\phi\lb z\rb=0,
  \eeq
  with
  \begin{subequations}
  \begin{gather}\label{alphas1}
  \alpha=1-\theta_0-\theta_1-\theta_t-\theta_\infty,\qquad \beta=1-\theta_0-\theta_1-\theta_t+\theta_\infty,\\
  \label{alphas2}
  \gamma=1-2\theta_0,\qquad \delta=1-2\theta_1,\qquad \epsilon=1-2\theta_t,\\
  \label{normcanc}
  q=\lb 1-t\rb\mathcal E+\frac{\gamma\epsilon+2\alpha\beta t-\lb\gamma+\delta\rb \epsilon t}{2}.
  \end{gather}
  \end{subequations}
  Obviously, we have a relation $\alpha+\beta+1=\gamma+\delta+\epsilon$.
  In spite of loosing a symmetry $\theta_k\mapsto -\theta_k$ ($k=0,1,t,\infty$) present in the normal form \eqref{heune}, the canonical form turns out to be better adapted for perturbative calculation below. A similar and very much related phenomenon is known from the AGT correspondence:  4D instanton partition functions have simpler series reperesentations than 2D conformal blocks but loose certain of their manifest symmetries because of the presence of the so-called $\mathrm{U}\lb 1\rb$ factor.

  Choose a basis in the space of solutions of \eqref{heunecan} which diagonalizes the composite monodromy around the pair of singular points $0$, $t$. Assuming that $0<|t|<1$,  its elements can be represented inside the annulus $|t|<|z|<1$ in the Floquet form
  \beq\label{Floqform}
  \phi\lb z\rb=\sum_{n\in \mathbb Z}c_n z^{n+\omega}.
  \eeq
  Substituting this series into \eqref{heunecan}, one obtains a linear 3-term recurrence relation  for the coefficients,
    \beq\label{3term}
  A_n c_{n-1}-B_n c_n+C_n t c_{n+1}=0,
  \eeq
  where
  \begin{subequations}
  \begin{align}
  A_n=&\,\lb \omega+n-1+\alpha\rb \lb \omega+n-1+\beta\rb,\\
  \label{coefsBn} B_n=&\,q+\lb \omega+n\rb \lb \epsilon+\delta t+\lb\omega+n-1+\gamma\rb\lb1+t\rb\rb,\\
  C_n=&\,\lb \omega+n+1\rb\lb\omega+n+\gamma\rb.
  \end{align}
  \end{subequations}
  The recurrence relation above is the main reason we started with \eqref{heunecan} instead of \eqref{heune}. As we will see in a moment, the crucial point for the computation of the small $t$ expansion of the accessory parameter,
  \beq\label{qexp}
  q\lb t\rb=\sum_{n=0}^{\infty}q_nt^n,
  \eeq
  is how the time $t$ appears in \eqref{3term}. 
  
  For $n\ge0$, define $u_n=\ds\frac{c_{n+1}}{c_n}$. This change of variables transforms \eqref{3term} into a nonlinear 2-term ``Riccati'' equation $u_{n-1}=\frac{A_n}{B_n-tC_n u_n}$. Under assumption that $B_{n}=O\lb 1\rb$ as $t\to0$ for $n\ge1$, one can then express $u_0$ as an infinite continued fraction
  \beq
  u_0=\frac{A_1}{B_1-\frac{t C_1A_2}{B_2-\frac{tC_2A_3}{B_3-\ldots}}}.
  \eeq
  For $n\leq0$, we first define rescaled coefficients $d_{-n}=c_nt^{n}$ and rewrite the $3$-term relation \eqref{3term} as
  \beq\label{3termII}
  tA_{-n} d_{n+1}-B_{-n} d_n+C_{-n} d_{n-1}=0,
  \eeq
  Introducing $v_n=\ds\frac{d_{n+1}}{d_n}$, we have $v_{n-1}= \frac{C_{-n}}{B_{-n}-tA_{-n} v_n}$, so that
  \beq
  v_0=\frac{C_{-1}}{B_{-1}-\frac{tA_{-1}C_{-2}}{B_{-2}-\frac{tA_{-2}C_{-3}}{B_{-3}-\ldots}}},
  \eeq
  under assumption that $B_n=O\lb 1\rb$ as $t\to 0$ for $n\leq -1$.
  From  $A_0c_{-1}-B_0c_0+tC_0c_1=0$ it follows that $t\lb C_0u_0+A_0v_0\rb=B_0$, which ultimately gives an equation determining the accessory parameter $q$ as a function of $t$ for given $\omega$: 
  \beq\label{contfraceq}
  \quad\frac{tC_0A_1}{B_1-\frac{t C_1A_2}{B_2-\frac{tC_2A_3}{B_3-\ldots}}}+\frac{tA_0C_{-1}}{B_{-1}-\frac{tA_{-1}C_{-2}}{B_{-2}-\frac{tA_{-2}C_{-3}}{B_{-3}-\ldots}}}=B_0.
  \eeq
  Note that $\left\{A_n\right\}$ and $\left\{C_n\right\}$ are just some monodromy-dependent constants; thus $q$ enters into \eqref{contfraceq} only through $\left\{B_n\right\}$.
  Substituting into \eqref{coefsBn} the expansion \eqref{qexp}, the coefficients $q_k$ can be recursively determined from \eqref{contfraceq} by truncating the continued fraction ladder at the desired order in $t$. 
  
  For example, at order $O\lb 1\rb$ we have $B_0=O\lb t\rb$, and therefore
  \begin{subequations}
  \beq
  q_0=-\omega\lb \omega+ \gamma+\epsilon-1\rb.
  \eeq
  At order $O\lb t\rb$, one has $B_0=t\lb\frac{C_0A_1}{B_1}+\frac{A_0C_{-1}}{B_{-1}}\rb+O\lb t^2\rb$, which gives
  \beq
  q_1=-\omega\lb \omega +\gamma+\epsilon-1\rb+\frac{\lb\omega+1\rb\lb\omega+\alpha\rb\lb\omega+\beta\rb\lb\omega+\gamma\rb}{2\omega+\gamma+\epsilon}-\frac{
  	\omega\lb\omega+\alpha-1\rb\lb\omega+\beta-1\rb\lb\omega+\gamma-1\rb}{2\omega+\gamma+\epsilon-2},
  \eeq
  \end{subequations}
  and so on. Substituting these coefficients into \eqref{normcanc}, we obtain the expansion of $\mathcal E=\mathcal E^{[\mathrm{F}]}_{\mathrm{VI}}\lb t|\,\sigma\rb$:
  \beq\label{heunAPexp}
  \begin{aligned}
  &\,\mathcal E^{[\mathrm{F}]}_{\mathrm{VI}}\lb t|\,\sigma\rb=\lb \delta_\sigma-\delta_0-\delta_t\rb+\mathcal{W}_1\lb \left\{\delta_k\right\}\rb t + 2\mathcal{W}_2\lb \left\{\delta_k\right\}\rb t^2+O\lb t^3\rb,
  \end{aligned}
  \eeq
  where $\mathcal{W}_{1,2}\lb \left\{\delta_k\right\}\rb$ are given by \eqref{W1expr}--\eqref{W2expr}, $\left\{\delta_k\right\}$ are defined by \eqref{scdims},  $\left\{\theta_k\right\}$ are related to $\alpha,\beta,\gamma,\delta,\epsilon$ by \eqref{alphas1}--\eqref{alphas2}, and the Floquet exponent appears only in ${\sigma=\omega-\theta_0-\theta_t+\frac12}$. One thus easily recognizes in \eqref{heunAPexp} the expansion of the logarithmic derivative $t\frac{\partial}{\partial t}\mathcal{W}\lb t\,\bigl|\left\{\delta_k\right\}\rb$ of the classical regular conformal block \eqref{clcblock}.

   	\section{Floquet characteristics for confluent Heun equations\label{appCH}}
    Throughout this section we refer to the notations of Table \ref{Table1}. The perturbative computation of accessory parameter functions of Floquet type presented here allows to check confluent Conjecture B \eqref{aux1823} at any desired order in $t$. Assuming the conjecture is true, this technique provides the most elementary method of computing classical CBs of the 1st kind.
    \subsection{Equation H$_\mathrm{V}$}
    The change of variables $\psi\lb z\rb=z^{\frac12-\theta_0}\lb z-t\rb^{\frac12-\theta_t}e^{\frac{z}{2}}\phi\lb z\rb$ transforms the confluent HE into the canonical form,
     \beq\label{heunPVcan}
   \phi''\lb z\rb+\lb\frac{\beta}{z}+\frac{\gamma}{z-t}+1\rb\phi'\lb z\rb+\frac{\alpha z-q}{z\lb z-t\rb}\phi\lb z\rb=0,
   \eeq
   with 
    \begin{subequations}
   	\begin{gather}\label{alphasHV}
   	\alpha=1-\theta_0-\theta_t-\theta_*,\qquad
   	\beta=1-2\theta_0,\qquad \gamma=1-2\theta_t,\\
   	\label{normcancHV}
   	q=-\mathcal E+\alpha t-\tfrac12{\lb\beta +t\rb\gamma}.
   	\end{gather}
   \end{subequations}
   Looking for the solutions of \eqref{heunPVcan} in the Floquet form \eqref{Floqform}, we arrive at the same continued fraction equation \eqref{contfraceq} for $q\lb t\rb$, except that the coefficients $\{A_n\}$, $\{B_n\}$, $\{C_n\}$ are now given by
     \begin{subequations}
   	\begin{align}
   	A_n=&\, \omega+n-1+\alpha ,\\
   	\label{coefsBnHV} B_n=&\,q-\lb \omega+n\rb \lb \omega+n-1+\beta+\gamma-t\rb,\\
   	C_n=&\,-\lb \omega+n+1\rb\lb\omega+n+\beta\rb.
   	\end{align}
   \end{subequations}
    The expansion of $q\lb t\rb$ can now be computed to any desired order. Its first terms read
    \beq
    q\lb t\rb=\omega\lb\omega+\beta+\gamma-1\rb+\left[-\omega+\frac{\lb\omega+1\rb\lb \omega+\alpha\rb\lb \omega+\beta\rb}{2\omega+\beta+\gamma}-\frac{\omega\lb \omega+\alpha-1\rb\lb \omega+\beta-1\rb}{2\omega+\beta+\gamma-2}\right]t+O\lb t^2\rb.
    \eeq
    The expansion of the Floquet characteristic $\mathcal E$ is then obtained from \eqref{normcancHV} (note that, in contrast with non-confluent HE, the coefficients of expansions of $\mathcal E$ and $-q$ coincide starting from the quadratic term). If we denote $\sigma=\omega-\theta_0-\theta_t+\frac12$ as before, then
    \beq
    \begin{aligned}
    \mathcal E^{[\mathrm{F}]}_{\mathrm{V}}\lb t|\,\sigma\rb=&\,\lb \delta_\sigma-\delta_0-\delta_t\rb-\frac{\theta_*\lb \delta_\sigma-\delta_0+\delta_t\rb}{2\delta_\sigma} t+\\
    +&\,\left[\frac{\theta_*^2\lb\delta_\sigma^2-\lb \delta_0-\delta_t\rb^2\rb}{8\delta_\sigma^3}-\frac{\lb 3\theta_*^2+\delta_\sigma\rb\lb \delta_\sigma^2+2\delta_\sigma\lb\delta_0+\delta_t\rb-3\lb\delta_0-\delta_t\rb^2\rb}{8\delta_\sigma^2\lb 3+4\delta_\sigma\rb}\right]t^2+O\lb t^3\rb.
    \end{aligned}
    \eeq
    It is instructive to check that this indeed agrees with the limit \eqref{conflimit1} and reproduces the expansion of
    $t\frac{\partial}{\partial t}\mathcal W_{N_f=3}\lb t\rb$, cf \eqref{NSpNf3}.

      \subsection{Equation H$_{\mathrm{III}_1}$}
      After the transformation $\psi\lb z\rb=\sqrt{z}\,e^{\frac{z}{2}+\frac{t}{2z}}\phi\lb z\rb$, the equation H$_{\mathrm{III}_1}$  becomes
      \beq\label{pseudocanHIII}
      \phi''\lb z\rb+\lb 1+\frac1z-\frac{t}{z^2}\rb\phi'\lb z\rb+\lb\frac{t\lb \frac12-\theta_\star\rb}{z^3}+\frac{\mathcal E-\frac t2-\frac14}{z^2}+\frac{\frac12-\theta_*}{z}\rb \phi\lb z\rb=0.
      \eeq
      This is \textit{not} the canonical form of doubly confluent Heun equation given in \href{http://dlmf.nist.gov/31.12.E2}{http://dlmf.nist.gov/31.12.E2}, which is obtained by a slightly different change of variables; yet it is more convenient for us to continue with \eqref{pseudocanHIII}. The Floquet substitution \eqref{Floqform} with $\omega=\sigma$ yields the equation \eqref{contfraceq} with
      	\beq
      	A_n=\theta_*-\sigma-n+\tfrac12 ,\qquad
      	\label{coefsBnHIII} B_n=\mathcal E-\tfrac t2-\tfrac14+\lb \sigma+n\rb^2 ,\qquad
      	C_n=\theta_\star+\sigma+n+\tfrac12,
      	\eeq
	 which then allows to compute the small $t$ expansion of $\mathcal E$. Its first few terms are given by
	   \beq
	 	\mathcal E^{[\mathrm{F}]}_{\mathrm{III}_1}\lb t|\,\sigma\rb= \delta_\sigma+\frac{\theta_*\theta_\star}{2\delta_\sigma} t+\lb \frac{\lb 3\theta_*^2+\delta_\sigma\rb \lb 3\theta_\star^2+\delta_\sigma\rb}{8\delta_\sigma^2\lb 3+4\delta_\sigma\rb}-\frac{\theta_*^2\theta_\star^2}{8\delta_\sigma^3}\rb t^2+O\lb t^3\rb.
	 \eeq
	
	  \subsection{Equation H$_{\mathrm{III}_2}$} 
	  In this case, the relevant change of variables is $\psi\lb z\rb=\sqrt{z}\,e^{\frac{z}{2}}\phi\lb z\rb$. It transforms the equation H$_{\mathrm{III}_2}$  into
	  \beq
	  \label{pseudocanHIII2}
	  \phi''\lb z\rb+\lb 1+\frac1z\rb\phi'\lb z\rb+\lb-\frac{t}{z^3}+\frac{\mathcal E-\frac14}{z^2}+\frac{\frac12-\theta_*}{z}\rb \phi\lb z\rb=0.
	  \eeq
	  The Floquet substitution \eqref{Floqform} with $\omega=\sigma$ again gives the equation \eqref{contfraceq}, whose coefficients are now given by
	  \beq
	  A_n=\theta_*-\sigma-n+\tfrac12 ,\qquad
	  \label{coefsBnHIII2} B_n=\mathcal E-\tfrac14+\lb \sigma+n\rb^2 ,\qquad
	  C_n=1,
	  \eeq
	  The expansion of accessory parameter function reads
	  \beq
	  \mathcal E^{[\mathrm{F}]}_{\mathrm{III}_2}\lb t|\,\sigma\rb= \delta_\sigma+\frac{\theta_*}{2\delta_\sigma} t+\frac{\lb 5\delta_\sigma-3\rb\theta_*^2+3\delta_\sigma^2}{8\delta_\sigma^3\lb 3+4\delta_\sigma\rb} t^2+\frac{\theta_*\lb\lb7\delta_\sigma-6\rb\delta_\sigma^2+\lb 9\delta_\sigma^2-19\delta_\sigma+6\rb\theta_*^2\rb}{16\delta_\sigma^5\lb 3+4\delta_\sigma\rb\lb 2+\delta_\sigma\rb}t^3+O\lb t^4\rb.
	  \eeq
	  
	    \subsection{Equation H$_{\mathrm{III}_3}$}
	   In this last case (recall that it is equivalent to Mathieu equation) the analog of the above calculations is rather well-known. There is no need to transform the equation. Making the Floquet ansatz $\psi\lb z\rb=\sum\limits_{n\in\mathbb Z}c_n z^{\frac12+\sigma+n}$ directly in the normal form, we arrive at a 3-term recurrence relation and equation \eqref{contfraceq} for~$\mathcal E$ with 
	  	  \beq
	  A_n=C_n=1,\qquad
	  \label{coefsBnHIII3} B_n=\mathcal E-\tfrac14+\lb \sigma+n\rb^2 ,
	  \eeq
	  This gives the expansion of the Mathieu characteristic value:
	  \beq
	  \mathcal E^{[\mathrm{F}]}_{\mathrm{III}_3}\lb t|\,\sigma\rb= \delta_\sigma+\frac{t}{2\delta_\sigma}+\frac{ 5\delta_\sigma-3}{8\delta_\sigma^3\lb 3+4\delta_\sigma\rb} t^2+\frac{ 9\delta_\sigma^2-19\delta_\sigma+6}{16\delta_\sigma^5\lb 3+4\delta_\sigma\rb\lb 2+\delta_\sigma\rb}t^3+O\lb t^4\rb.
	  \eeq
	  To facilitate comparison with the literature, note that the parameters $\nu$ and $q$ in e.g. \cite[Eq. 20.3.15]{AS} and \cite[\href{http://dlmf.nist.gov/28.15.E1}{Eq. 28.15.E1}]{DLMF} correspond to our $2\sigma$ and $4\sqrt t$ so that $\delta_\sigma=\frac{1-\nu^2}{4}$.

\end{document}

%% file: diagram.tex
\begin{tikzpicture}[scale=0.3]

\node (H) at (1.2, -6){H$_{\mathrm{VI}}$};

\coordinate (H) at (-1.5,1.5);
\coordinate (e1) at (H);
\draw[rotate around={45:(e1)}] (e1) ellipse (1.41 and 0.7);
\draw ($(e1)+(1,1)$) .. controls +(0.6,-0.8) and +(-0.6,-0.8)  .. +(3,0);
\coordinate (e2) at ($(e1)+(5,0)$);
\draw[rotate around={-45:(e2)}] (e2) ellipse (1.41cm and 0.7cm);
\draw ($(e2)+(1,-1)$) .. controls +(-0.8,-0.6) and +(-0.8,0.6)  .. +(0,-3);
\coordinate (e3) at ($(e1)+(5,-5)$);
\draw[rotate around={45:(e3)}] (e3) +(180:1.41) arc (180:360:1.41 and 0.7);
\draw[rotate around={45:(e3)},dashed,opacity=0.6] (e3) +(0:1.41) arc (0:180:1.41 and 0.7);
\draw ($(e3)+(-1,-1)$) .. controls +(-0.6,0.8) and +(0.6,0.8)  .. +(-3,0);
\coordinate (e4) at ($(e1)+(0,-5)$);
\draw[rotate around={-45:(e3)}] (e4) +(180:1.41) arc (180:360:1.41 and 0.7);
\draw[rotate around={-45:(e3)},dashed,opacity=0.6] (e4) +(0:1.41) arc (0:180:1.41 and 0.7);
\draw ($(e4)+(-1,1)$) .. controls +(0.8,0.6) and +(0.8,-0.6)  .. +(0,3);
\draw[dashed,opacity=0.6,red] ($(e1)+(2.5,-2.5)$) +(0:2.9) arc (0:180:2.9 and 0.5);
\draw[red] ($(e1)+(2.5,-2.5)$) +(0:2.9) arc (0:-180:2.9 and 0.5);
\node (a) at (e1){$\infty$};
\node (b) at (e2){$1$};
\node (c) at (e3){$t$};
\node (d) at (e4){$0$};

\draw[->,thick] (5,-1)--(11,-1);

\node (CH) at (12, -5){H$_{\mathrm V}$};

\coordinate (CH) at (15,1);
\coordinate (e1) at (CH);
\draw (e1) +(180:2) arc (180:90:2 and 0.7);
\draw ($(e1)+(0,0.7)$) .. controls +(0.8,0)  .. +(1,0.8);
\draw ($(e1)+(2,0)$) .. controls +(0,0.8) and +(0,-1.3)   .. +(-1,1.5);
\draw (e1) +(0:2) arc (0:-90:2 and 0.7);
\draw ($(e1)+(0,-0.7)$) .. controls +(-0.5,0) and +(0,-0.7)  .. +(-1,0.9);
\draw ($(e1)+(-2,0)$) .. controls +(0,-0.5) and +(0,-1)   .. +(1,0.2);
\draw ($(e1)+(2,0)$) .. controls +(0,-1)  .. +(1.5,-2.5);
\draw ($(e1)+(-2,0)$) .. controls +(0,-1)  .. +(-1.5,-2.5);
\coordinate (e3) at ($(e1)+(2.5,-3.5)$);
\draw[rotate around={45:(e3)}] (e3) +(180:1.41) arc (180:360:1.41 and 0.7);
\draw[rotate around={45:(e3)},dashed,opacity=0.6] (e3) +(0:1.41) arc (0:180:1.41 and 0.7);
\draw ($(e3)+(-1,-1)$) .. controls +(-0.6,0.8) and +(0.6,0.8)  .. +(-3,0);
\coordinate (e4) at ($(e3)+(-5,0)$);
\draw[rotate around={-45:(e3)}] (e4) +(180:1.41) arc (180:360:1.41 and 0.7);
\draw[rotate around={-45:(e3)},dashed,opacity=0.6] (e4) +(0:1.41) arc (0:180:1.41 and 0.7);
\draw[dashed,opacity=0.6,red] ($(e1)+(0,-1.5)$) +(0:2.5) arc (0:180:2.5 and 0.5);
\draw[red] ($(e1)+(0,-1.5)$) +(0:2.5) arc (0:-180:2.5 and 0.5);

\draw[->,thick] (18,0)--(22,2.5);

\node (RCH) at (22, 7){H$_{\mathrm{III}_1^\prime}$};

\coordinate (RCH) at (26,6);
\coordinate (e1) at (RCH);
\draw (e1) +(180:2) arc (180:0:2 and 0.7);
\draw (e1) +(0:2) arc (0:-90:2 and 0.7);
\draw ($(e1)+(0,-0.7)$) .. controls +(-0.5,0) and +(0,-0.7)  .. +(-1,0.9);
\draw ($(e1)+(-2,0)$) .. controls +(0,-0.5) and +(0,-1)   .. +(1,0.2);
\draw ($(e1)+(2,0)$) .. controls +(0,-1)  .. +(1.5,-2.5);
\draw ($(e1)+(-2,0)$) .. controls +(0,-1)  .. +(-1.5,-2.5);
\coordinate (e3) at ($(e1)+(2.5,-3.5)$);
\draw[rotate around={45:(e3)}] (e3) +(180:1.41) arc (180:360:1.41 and 0.7);
\draw[rotate around={45:(e3)},dashed,opacity=0.6] (e3) +(0:1.41) arc (0:180:1.41 and 0.7);
\draw ($(e3)+(-1,-1)$) .. controls +(-0.6,0.8) and +(0.6,0.8)  .. +(-3,0);
\coordinate (e4) at ($(e3)+(-5,0)$);
\draw[rotate around={-45:(e3)}] (e4) +(180:1.41) arc (180:360:1.41 and 0.7);
\draw[rotate around={-45:(e3)},dashed,opacity=0.6] (e4) +(0:1.41) arc (0:180:1.41 and 0.7);
\draw[dashed,opacity=0.6,red] ($(e1)+(0,-1.5)$) +(0:2.5) arc (0:180:2.5 and 0.5);
\draw[red] ($(e1)+(0,-1.5)$) +(0:2.5) arc (0:-180:2.5 and 0.5);

\draw[->,thick] (20,-3.5)--(29,-7.5);

\node (DCH) at (31, -4){H$_{\mathrm{III}_1}$};

\coordinate (DCH) at (32,-6);
\coordinate (e1) at (DCH);
\draw (e1) +(180:2) arc (180:90:2 and 0.7);
\draw ($(e1)+(0,0.7)$) .. controls +(0.8,0)  .. +(1,0.8);
\draw ($(e1)+(2,0)$) .. controls +(0,0.8) and +(0,-1.3)   .. +(-1,1.5);
\draw (e1) +(0:2) arc (0:-90:2 and 0.7);
\draw ($(e1)+(0,-0.7)$) .. controls +(-0.5,0) and +(0,-0.7)  .. +(-1,0.9);
\draw ($(e1)+(-2,0)$) .. controls +(0,-0.5) and +(0,-1)   .. +(1,0.2);
\draw ($(e1)+(2,0)$) -- +(0,-4);
\draw ($(e1)+(-2,0)$) -- +(0,-4);
\coordinate (e2) at ($(e1)+(0,-4)$);
\draw[densely dashed,opacity=0.6] (e2) +(180:2) arc (180:90:2 and 0.7);
\draw[densely dashed,opacity=0.6] ($(e2)+(0,0.7)$) .. controls +(0.8,0)  .. +(1,-0.5);
\draw[densely dashed,opacity=0.6] ($(e2)+(2,0)$) .. controls +(0,0.5) and +(0,0.6)   .. +(-1,0.2);
\draw (e2) +(0:2) arc (0:-90:2 and 0.7);
\draw ($(e2)+(0,-0.7)$) .. controls +(-0.5,0) and +(0,0.3)  .. +(-1,-0.5);
\draw ($(e2)+(-2,0)$) .. controls +(0,-0.5) and +(0,0.7)   .. +(1,-1.2);
\draw[dashed,opacity=0.6,red] ($(e1)+(0,-2)$) +(0:2) arc (0:180:2 and 0.5);
\draw[red] ($(e1)+(0,-2)$) +(0:2) arc (0:-180:2 and 0.5);

\draw[->,thick] (30,2)--(37,-1.2);
\draw[->,thick] (35,-7.5)--(38,-5);

\node (RDCH) at (42, 2){H$_{\mathrm{III}_2}$};

\coordinate (RDCH) at (40,-2);
\coordinate (e1) at ($(RDCH)+(0,2)$);
\draw (e1) +(180:2) arc (180:0:2 and 0.7);
\draw (e1) +(0:2) arc (0:-90:2 and 0.7);
\draw ($(e1)+(0,-0.7)$) .. controls +(-0.5,0) and +(0,-0.7)  .. +(-1,0.9);
\draw ($(e1)+(-2,0)$) .. controls +(0,-0.5) and +(0,-1)   .. +(1,0.2);
\draw ($(e1)+(2,-0.1)$) -- +(0,-3.8);
\draw ($(e1)+(-2,-0.1)$) -- +(0,-3.8);
\coordinate (e2) at ($(e1)+(0,-4)$);
\draw[densely dashed,opacity=0.6] (e2) +(180:2) arc (180:90:2 and 0.7);
\draw[densely dashed,opacity=0.6] ($(e2)+(0,0.7)$) .. controls +(0.8,0)  .. +(1,-0.8);
\draw[densely dashed,opacity=0.6] ($(e2)+(2,0)$) .. controls +(0,0.5) and +(0,0.6)   .. +(-1,0);
\draw (e2) +(0:2) arc (0:-90:2 and 0.7);
\draw ($(e2)+(0,-0.7)$) .. controls +(-0.5,0) and +(0,0.6)  .. +(-1,-0.8);
\draw ($(e2)+(-2,0)$) .. controls +(0,-0.5) and +(0,1)   .. +(1,-1.4);
\draw[dashed,opacity=0.6,red] ($(e1)+(0,-2)$) +(0:2) arc (0:180:2 and 0.5);
\draw[red] ($(e1)+(0,-2)$) +(0:2) arc (0:-180:2 and 0.5);

\draw[->,thick] (43,-2)--(50,-2);

\node (DRDCH) at (52, -7){H$_{\mathrm{III}_3}$};

\coordinate (DRDCH) at (53,-2);
\coordinate (e1) at ($(DRDCH)+(0,2)$);
\draw (e1) +(180:2) arc (180:0:2 and 0.7);
\draw (e1) +(0:2) arc (0:-90:2 and 0.7);
\draw ($(e1)+(0,-0.7)$) .. controls +(-0.5,0) and +(0,-0.7)  .. +(-1,0.9);
\draw ($(e1)+(-2,0)$) .. controls +(0,-0.5) and +(0,-1)   .. +(1,0.2);
\draw ($(e1)+(2,-0.1)$) -- +(0,-3.8);
\draw ($(e1)+(-2,-0.1)$) -- +(0,-3.8);
\coordinate (e2) at ($(e1)+(0,-4)$);
\draw[densely dashed,opacity=0.6] (e2) +(180:2) arc (180:0:2 and 0.7);
\draw (e2) +(0:2) arc (0:-90:2 and 0.7);
\draw ($(e2)+(0,-0.7)$) .. controls +(-0.5,0) and +(0,0.6)  .. +(-1,-0.8);
\draw ($(e2)+(-2,0)$) .. controls +(0,-0.5) and +(0,1)   .. +(1,-1.4);
\draw[dashed,opacity=0.6,red] ($(e1)+(0,-2)$) +(0:2) arc (0:180:2 and 0.5);
\draw[red] ($(e1)+(0,-2)$) +(0:2) arc (0:-180:2 and 0.5);

\draw[->,thick] (15,-5)--(15,-13);

\node (BE) at (12, -14){H$_{\mathrm{IV}}$};

\coordinate (BE) at (15,-18);
\coordinate (e1) at ($(BE)+(0,2)$);
\coordinate (e2) at ($(e1)+(0,-3)$);
\draw (e1) +(180:3) arc (180:120:3 and 0.7);
\draw ($(e1)+(-1.5,0.6)$) .. controls +(0.5,0) and +(-0.1,-0.7)  .. +(1.5,0.8);
\draw ($(e1)+(0,1.4)$) .. controls +(0.1,-0.8) and +(-0.4,-0.8)  .. +(1.3,-0.3);
\draw ($(e1)+(3,0)$) .. controls +(0,0.5) and +(0,-1)   .. +(-1.7,1.2);
\draw (e1) +(0:3) arc (0:-60:3 and 0.7);
\draw ($(e1)+(1.5,-0.6)$) .. controls +(-0.5,0) and +(0,-1.2)  .. +(-1.5,0.8);
\draw ($(e1)+(0,0.2)$) .. controls +(-0.2,-1.1) and +(-0.1,-1.1)   .. +(-1.3,0.1);
\draw ($(e1)+(-3,0)$) .. controls +(0,-0.7) and +(0,-1)   .. +(1.7,0.3);
\draw ($(e2)+(2,0)$) .. controls +(0,2) and +(0,-2) .. ($(e1)+(3,0)$);
\draw ($(e2)+(-2,0)$) .. controls +(0,2) and +(0,-2) .. ($(e1)+(-3,0)$);
\draw[densely dashed,opacity=0.6] (e2) +(180:2) arc (180:0:2 and 0.7);
\draw (e2) +(0:2) arc (0:-180:2 and 0.7);

\draw[thick] (26,1)--(26,-5.7);
\draw[->, thick] (26,-6.7)--(26,-9);
\draw[->,thick] (19,-15)--(22.5,-12.5);

\node (RBHE) at (26, -16){H$_{\mathrm{II}^\prime}$};

\coordinate (RBHE) at (26,-13);
\coordinate (e1) at ($(RBHE)+(0,2)$);
\coordinate (e2) at ($(e1)+(0,-3)$);
\draw (e1) +(180:3) arc (180:90:3 and 0.7);
\draw ($(e1)+(0,0.7)$) .. controls +(0.6,0) and +(-0.2,-0.6)  .. +(1.3,0.7);
\draw ($(e1)+(3,0)$) .. controls +(0,0.7) and +(0,-1.2)   .. +(-1.7,1.4);
\draw (e1) +(0:3) arc (0:-60:3 and 0.7);
\draw ($(e1)+(1.5,-0.6)$) .. controls +(-0.5,0) and +(0,-1.2)  .. +(-1.5,0.8);
\draw ($(e1)+(0,0.2)$) .. controls +(-0.2,-1.1) and +(-0.1,-1.1)   .. +(-1.3,0.1);
\draw ($(e1)+(-3,0)$) .. controls +(0,-0.7) and +(0,-1)   .. +(1.7,0.3);
\draw ($(e2)+(2,0)$) .. controls +(0,2) and +(0,-2) .. ($(e1)+(3,0)$);
\draw ($(e2)+(-2,0)$) .. controls +(0,2) and +(0,-2) .. ($(e1)+(-3,0)$);
\draw[densely dashed,opacity=0.6] (e2) +(180:2) arc (180:0:2 and 0.7);
\draw (e2) +(0:2) arc (0:-180:2 and 0.7);

\draw[->, thick] (32,-12)--(32,-21);
\draw[->,thick] (20,-19)--(28,-23);

\node (TRI) at (30, -21){H$_{\mathrm{II}}$};

\coordinate (TRI) at (32,-25);
\coordinate (e1) at ($(TRI)+(0,2)$);
\draw ($(e1)+(-3,0)$) .. controls +(0,0.5) and +(0,-1)   .. +(1.7,1.2);
\draw ($(e1)+(-1.3,1.2)$) .. controls +(0,-1) and +(0,-0.8)  .. +(1.3,0.1);
\draw ($(e1)+(0,1.2)$) .. controls +(0.2,-0.6) and +(-0.4,-0.8)  .. +(1.3,-0.1);
\draw ($(e1)+(3,0)$) .. controls +(0,0.5) and +(0,-1)   .. +(-1.7,1.2);

\draw ($(e1)+(3,0)$) .. controls +(0,-0.7) and +(0,-1)   .. +(-1.7,0.3);
\draw ($(e1)+(1.3,0.3)$) .. controls +(0,-1) and +(0,-1.2)  .. +(-1.3,-0.1);
\draw ($(e1)+(0,0.2)$) .. controls +(-0.2,-1.1) and +(-0.1,-1.1)   .. +(-1.3,0.1);
\draw ($(e1)+(-3,0)$) .. controls +(0,-0.7) and +(0,-1)   .. +(1.7,0.3);
\draw (e1) +(0:3) arc (0:-180:3 and 3);

\draw[thick] (30,-13)--(31.5,-13.8);
\draw[->, thick] (32.5,-14.2)--(36,-16);
\draw[->,thick] (34,-22)--(37,-19.5);
\draw[->,thick] (40,-7)--(40,-14);

\node (RTRI) at (43, -15){H$_{\mathrm I}$};
q
\coordinate (RTRI) at (40,-19);
\coordinate (e1) at ($(RTRI)+(0,2)$);
\draw (e1) +(180:3) arc (180:120:3 and 0.7);
\draw ($(e1)+(-1.5,0.6)$) .. controls +(1,0) and +(0,-0.6)  .. +(1.5,0.8);
\draw ($(e1)+(0,1.4)$) .. controls +(0.2,-0.8) and +(-0.4,-0.8)  .. +(1.3,-0.2);
\draw ($(e1)+(3,0)$) .. controls +(0,0.5) and +(0,-1)   .. +(-1.7,1.2);
\draw ($(e1)+(3,0)$) .. controls +(0,-0.7) and +(0,-1)   .. +(-1.7,0.3);
\draw ($(e1)+(1.3,0.3)$) .. controls +(0,-1) and +(0,-1.2)  .. +(-1.3,-0.1);
\draw ($(e1)+(0,0.2)$) .. controls +(-0.2,-1.1) and +(-0.1,-1.1)   .. +(-1.3,0.1);
\draw ($(e1)+(-3,0)$) .. controls +(0,-0.7) and +(0,-1)   .. +(1.7,0.3);
\draw (e1) +(0:3) arc (0:-180:3 and 3);

\end{tikzpicture}

%% file: 4point.tex
\begin{tikzpicture}[scale=0.7]

\draw[thick] (0,0) -- (2,0);
\draw[thick] (2,0) -- (2,2);
\draw[thick, red] (2,0) -- (4,0);
\draw[thick] (4,0) -- (4,2);
\draw[thick] (4,0) -- (6,0);

\node () at (-0.5, 0){$\infty$};
\node () at (2, 2.5){$1$};
\node () at (4, 2.5){$t$};
\node () at (6.5, 0){$0$};

\node () at (1.5, 1.5){$\Delta_1$};
\node () at (4.5, 1.5){$\Delta_t$};
\node () at (0.5, -0.5){$\Delta_\infty$};
\node[red] () at (3, -0.5){$\Delta_\sigma$};
\node () at (5.5, -0.5){$\Delta_0$};
\end{tikzpicture}

%% file: conflBlock1.tex
\begin{tikzpicture}[scale=0.7]

\draw[thick] (0,-0.05) -- (2,-0.05);
\draw[thick] (0,0.05) -- (2,0.05);
\draw[thick, red] (2,0) -- (4,0);
\node[circle, fill, inner sep=1pt] () at (2,0){};
\draw[thick] (4,0) -- (4,2);
\draw[thick] (4,0) -- (6,0);

\node () at (-0.5, 0){$\infty$};
\node () at (4, 2.5){$t$};
\node () at (6.5, 0){$0$};

\node () at (1, 0.5){$(P_*,\frac{1}{4})$};
\node () at (4.5, 1.5){$\Delta_t$};
\node[red] () at (3, -0.5){$\Delta_\sigma$};
\node () at (5.5, -0.5){$\Delta_0$};
\end{tikzpicture}

%% file: conflBlock2.tex
\begin{tikzpicture}[scale=0.7]

\draw[thick] (0,-0.05) -- (2,-0.05);
\draw[thick] (0,0.05) -- (2,0.05);
\draw[thick, red] (2,0.05) -- (4,0.05);
\draw[thick, red] (2,-0.05) -- (4,-0.05);
\node[circle, fill, inner sep=1pt] () at (2,0){};
\draw[thick] (2,0) -- (2,2);
\draw[thick] (4,0) -- (6,0);
\node[circle, fill, inner sep=1pt] () at (4,0){};

\node () at (-0.5, 0){$\infty$};
\node () at (2, 2.5){$t$};
\node () at (6.5, 0){$0$};

\node () at (1, -0.5){$(P_\ast,\frac{1}{4})$};
\node () at (2.5, 1.5){$\Delta_t$};
\node[red] () at (3, -0.5){$(P_\nu,\frac{1}{4})$};
\node () at (5.5, -0.5){$\Delta_0$};
\end{tikzpicture}

%% file: conflBlock3.tex
\begin{tikzpicture}[scale=1]

\draw[thick] (0,-0.1) -- (2,-0.1);
\draw[thick] (0,0) -- (2,0);
\draw[thick] (0,0.1) -- (2,0.1);
\draw[thick, red] (2,0.1) -- (4,0.1);
\draw[thick, red] (2,0) -- (4,0);
\draw[thick, red] (2,-0.1) -- (4,-0.1);
\node[circle, fill, inner sep=2pt] () at (2,0){};
\draw[thick] (2,0) -- (2,2);
\draw[thick, dashed] (4,0) -- (6,0);
\node[circle, fill, inner sep=2pt] () at (4,0){};

\node () at (-0.5, 0){$\infty$};
\node () at (2, 2.5){$t$};
\node () at (6.5, 0){$0$};

\node () at (1, -0.5){\small{$(P_\bullet, 0,\frac{1}{4})$}};
\node () at (2.5, 1.5){$\Delta_0$};
\node[red] () at (3, -0.5){\small{$(P_\bullet - P_\nu, 0 ,\frac{1}{4})$}};
\node () at (5.5, 0.5){$0$};
\end{tikzpicture}